# Functions of Bifans in Context of Multiple Regulatory Motifs in Signaling Networks

* Azi Lipshtat ,* Sudarshan P. Purushothaman ,* Ravi Iyengar ,* Avi Ma'ayan

* "Department of Pharmacology and Systems Therapeutics, Mount Sinai School of Medicine, New York, NY 10029, USA"

## ABSTRACT

Representation of intracellular signaling networks as directed graphs allows for the identification of regulatory motifs. Regulatory motifs are groups of nodes with the same connectivity structure, capable of processing information. The bifan motif, made of two source nodes directly cross-regulating two target nodes, is an over-represented motif in a mammalian cell signaling network and in transcriptional networks. One example of a bifan is the two MAP-kinases, p38 and JNK that phosphorylate and activate the two transcription factors ATF2 and Elk-1. We have used a system of coupled ordinary differential equations to analyze the regulatory capability of this bifan motif by itself, and when it interacts with other motifs such as positive and negative feedback loops. Our results indicate that bifans provide temporal regulation of signal propagation and act as signal sorters, filters, and synchronizers. Bifans that have OR gate configurations show rapid responses while AND gate bifans can introduce delays and allow prolongation of signal outputs. Bifans that are AND gates can filter noisy signal inputs. The p38/JNK-ATF2/Elk-1bifan synchronizes the output of activated transcription factors. Synchronization is a robust property of bifans and is exhibited even when the bifan is adjacent to a positive feedback loop. The presence of the bifan promotes the transcription and translation of the dual specificity protein phosphatase MKP-1 that inhibits p38 and JNK thus enabling a negative feedback loop. These results indicate that bifan motifs in cell signaling networks can contribute to signal processing capability both intrinsically and by enabling the functions of other regulatory motifs.

## INTRODUCTION

Systematic understanding of the design principles of regulatory circuits in cells is a necessary step towards building predictive models of mammalian cells (1, 2). Complex large-scale biochemical networks can be represented as graphs where nodes are molecular components and links represent their interactions (3, 4). Analysis of such networks allows for the identification of motifs, reoccurring sub-graphs of few interacting nodes (5-7). In cell signaling networks and gene regulatory networks, regulatory motifs such as feed-forward (5, 8-10), feedback loops (11, 12), single-input modules (SIM) (13, 14) and bifans (9, 11, 15), represent small circuits with information processing capabilities that can alter input/output relationships within the network. Quantitative analysis of these motifs can be used to identify the information processing capabilities of these circuits. Several studies characterizing the dynamical behavior of network motifs such as feed-forward (8, 16, 17), feedback loops (18-21), and bifans (22), as well as stability analysis of different motifs (23) have been published. These studies reported that feed forward motifs with AND configuration can exhibit a delay following by signal activation (8, 13, 16, 24) and filter noise (17, 25), while feed forward motifs with SUM configuration show delay after signal deactivation (10). Feedback loops can lead to bistable behavior (18, 20, 21, 26-28), oscillations (12, 19, 29-32), signal delay (14, 33), and can filter noise (29). Single-input modules (SIMs) (18) are sets of operons that are controlled by a single transcription factor that coordinates the activity of all the operons. This configuration can be used for signal sorting and output synchronization (13, 14) (See Table 1 for summary and comparison between the functional capabilities of the various motifs). Bifans, which have been less well studied, were shown to be highly dependent on the gating mechanisms used (i.e. AND or OR) (22). However, the functional characteristics of bifans have not been studied in depth. Since the bifan consists of multiple activation routes for each target, a reasonable hypothesis would be that bifans function is similar to those of the feed-forward loop motif. Stability analysis of all possible small-size motifs identified classes of motifs, where the bifan motif was found to belong to Class I of Structural Stability Score (SSS). Motifs in this class have no loops and are shown to be most stable, while motifs with feedback loops are are least stable (23). This analysis implied that bifan motifs are expected to have stable dynamics.

The bifan motif is the most statistically over-represented network motif in signaling (9, 11) and gene regulatory networks (5, 22). The configuration of the bifan motif includes two source nodes directly cross-regulating two target nodes (shaded area in Fig. 1a). When considering the type of nodes and type of links making up each bifan, the bifan motif consisting of two protein kinases each phosphorylating and activating two transcription factors was found to be highly over-represented in a mammalian neuronal intracellular signaling network we constructed from literature (11). Hence, in this study, we have investigated the dynamical properties of a cell signaling bifan network motif made of two protein kinases regulating the activity of two transcription factors. Using a system of coupled ODEs we determined the intrinsic capabilities of this bifan motif.

Mitogen-activated protein kinase p38α and c-Jun N-terminal kinase-1 (JNK) are two protein kinases that phosphorylate and activate the cAMP-dependent Activating transcription factor-2 (ATF2) and ETS domain protein Elk-1 transcription factors. Transcription factors, when phosphorylated by protein kinases, become activated, and often form homodimers and/or hetrodimers that participate in dynamic nuclear complexes (34) and bind DNA promoter sequences to regulate transcription. ATF2 belongs to a family of transcription factors that can homodimerize or heterodimerize with c-Fos, FosB, Fra1 and Fra2 or c-Jun, JunB, and JunD family members (35). As dimers, these transcription factors bind to specific promoter sites such as the AP1 consensus sequence TGAC/GTC/AA (36). When these dimers bind to the promoter they enable the assembly a complex that allows for DNA remodeling and initiation of transcription (37). Elk-1 forms homodimers and hetrodimers with TCF family members to bind to specific promoter elements such as SRF (38, 39). JNK1 and p38α are protein kinases that target and phosphorylate Elk-1 at two different domains (40) and are also known to phosphorylate ATF2 (41, 42) shown, for example, in response to UV irradiation and in neuroblastoma cells (43).

The bifan motif we have investigated in this study is fully coherent, where all links are positive (activating). A previous report on bifan motifs considered dynamics of bifans with both positive and negative links to analyze various possible architectures of gene regulatory networks (22), but it is not clear if these findings are applicable to cell signaling networks. Here we focus on a cell signaling bifan. We analyze this motif in depth, and show that even though this specific circuit can fit a large class of topological and dynamical possibilities, it is possible to deduce general novel conclusions about function and advantages of the bifan configuration. We identify parts of the dynamics that can be attributed to the motif itself, and parts that result from the surrounding connectivity environment.

The bifan motif of two protein kinases regulating two transcription factors does not function in isolation, rather such bifans are juxtaposed with other positive and negative feedback and forward motifs. For example, it is known that the transcription factor c-Jun is phosphorylated on serine 63 and serine 73 by JNK (44-47). Thus, JNK can phosphorylate Elk-1, ATF2 and c-Jun (48). ATF2 and c-Jun can form hetrodimers and c-Jun can also form homodimers (49). Hence, we analyzed the consequences of placing the bifan motif adjacent to the JNK/c-Jun interaction (shaded area in Fig. 1b). Additionally, downstream of the JNK/p38/ATF2/Elk-1 bifan motif there are immediate early genes (IEG) such as *c-Jun* and *MKP-1*. c-Jun homodimers can bind to the promoter site of the *c-Jun* gene, while at another site: the jun2 site, ATF2/c-Jun hetrodimers bind (50, 51). The binding of these homodimers and hetrodimers to these promoters induces *c-Jun* gene transcription, resulting in a positive feedback loop. There is evidence that the c-Jun protein is rapidly expressed after JNK and p38 activation reaching a maximum after 3 hours and with a slow decline after 12 hours (52). There is another known feedback loop in this system: the rapid induction of MKP-1. This is a negative feedback loop. MKP-1 transcription is induced by JNK and p38. MKP-1 is a dual specificity phosphatase that deactivates JNK and p38 (53-55). There is evidence that MKP-1 mRNA is induced in endothelial cells after TNF-α stimulation and the levels of MKP-1 mRNA increases after one hour. TNF-α is known to stimulate JNK and p38 and induce MKP-1 transcription (56). MKP-1 can dephosphorylate and deactivate p38, JNK and Erk1/2 MAPKs (57), while in U937 human leukemic cells MKP-1 was shown to be specific for JNK and p38 (54). In liver cells, oxidative stress induced JNK phosphorylation, and subsequent increase in AP1 activity involving activation of c-Jun, ATF2, c-Fos, JunB and JunD. MKP-1 is rapidly induced under oxidative stress in liver cells, and was shown to dephosphorylate p38 which is constitutively active under normal conditions (58). These observations suggest a complex hierarchy of interactions at the levels of protein-kinases, transcription-factor activation and IEG feedback, with MKP-1 playing the role of a negative feedback regulator. How might the presence of a bifan motif at the center of this topology influence the dynamics of this circuit? To understand the role of bifan motifs in this context, we analyzed the quantitative dynamical behavior of the bifan motif coupled to the positive feedback loop of c-Jun to itself (shaded area in Fig. 1c) and the negative feedback loop involving MKP-1 (shaded area in Fig. 1d). We used ordinary differential equation (ODE) simulations with a deterministic representation of the discrete nature of binding sites to deal with transcription regulation and protein translation. As far as we know, combining ODE models of a signaling pathway linked to a transcriptional circuit that feeds back to the signaling pathway, has not been implemented using such an approach before.

**RESULTS**

Although it is known that Elk-1 and ATF2 can be phosphorylated at several sites, it is not clear from the experimental studies whether the different protein kinases phosphorylate the same target sites and can activate the transcription factors independently (OR gate, Fig. 2a), or whether both protein kinases are required for activation (AND gate, Fig. 2b). For the AND gate case, it is possible to have a sequential hierarchy among the phosphorylations (i.e the AND gate is ordered such that the first phosphorylation is required to occur before the second phosphorylation). Hence, we modeled these three possibilities using rate constants obtained from literature (described in the Methods and in Tables 2 and 3). In principle, one could consider different gating mechanisms for each of the two transcription factors. Nevertheless, in this study we limited the analysis to have the same logical gate for both transcription factors. Comparison of the OR and AND gates indicates that the OR variant shows rapid response, whereas the AND gates cause delay and thus produce activated transcription factors at later times (Fig. 2 c-e). Activation of transcription factors (and thus dimer production) is maintained for some time even after the signal is turned off, due to the non-zero concentrations of p38* and JNK*. (Note that direct inactivation of TF is not included in this model. Active ATF2 is assumed to be removed from the system by dimerization.) When input signals were applied as a set of pulses of short duration, an initial short pulse was enough to start a significant production of ATF2 homodimers with the OR gate configuration (Fig. 2d). In contrast, multiple pulses were required to initiate significant activation in the AND gate configuration. With the ordered AND configuration, a short pulse activates the transcription factor to a level which is more than three fold lower than the activation level with the OR configuration (Fig. 2e). Hence, the dimerization rate, which is proportional to the square of the activated TF concentration, is about an order of magnitude slower in the ordered AND than in the OR configuration. These results suggest that AND gated bifans can function as signal delay resistors and coincidence detectors. These conclusions are similar to what was reported for AND and OR gating of feed forward motifs (8). As these studies were ongoing Ingram et al. reported similar findings (22). Our results indicate that it is necessary to specify the AND or OR gate topology of bifans to define their functional capabilities.

Some of the advantages of having two upstream effectors compared to just one are easily understood. For example, two activation routes can be used for redundancy in the OR configuration. If any single protein kinase is sufficient for full activation, the existence of an alternate signaling pathway provides more reliability. In the AND gate configuration, the benefit is a coincidence detector to guard against activation by a single input protein kinase. On the other hand, there are less obvious advantages as well. For example, there may be an indirect way by which the bifan configuration enables the presence of one transcription factor to effectively change the activation level of another transcription factor. We found that the initial concentration of one transcription factor (Elk-1) has a decisive effect on the activation rate of the other transcription factor (ATF2). These two transcription factors compete for the limited input from p38α and JNK1. Increasing the amount of Elk-1 causes a higher "trapping" of p38α and JNK by Elk-1 at early times (Fig. 3a), and thus decreasing the amounts of ATF2 that can be activated and in turn the ATF2 homodimers that can be produced (Fig. 3b). Thus, under defined conditions, Elk-1 can function as an inhibitory regulator of ATF2. This model does not include any direct ATF2 inactivation. Inclusion of such a reaction is expected to enhance this effect, since more p38α and JNK will be required for same ATF2 activation level. This effect, driven by different initial conditions, causes delay in dimer production and slows ATF2 activation as the initial amounts of Elk-1 increases. Thus, the concentration of one transcription factor can control and regulate the activation level of another transcription factor, a design feature that is important for gene expression regulation.

Changing environmental conditions as well as extrinsic stochastic fluctuations can affect the circuit dynamics (59, 60). To address this, we examined the response of the bifan motif to randomly generated input signals. When the bifan motif is stimulated with random input signals, an OR gate starts the activation process immediately, whereas the AND gate requires repeated or prolonged input signals for activation. The activation of transcription factors takes place only after the activation of sufficient levels of upstream activated components with the AND gating, allowing the bifan to filter out sporadic random input signals. Since the output of the AND gate depends on the duration of the input signal, the AND gate also serves as an integrator, responding to the total summation of signal over time, rather than responding to the signal at any given time point. Thus, even if the input (protein kinase activation) is noisy, the resultant output (levels of TFs dimers) is smooth (Fig. 4a). We also examined the output using only one protein kinase vs. activation of both protein kinases. It is noticeable that when only one of the protein kinases is functional, the output signal fluctuates more than when both protein kinases are present (AND gate). The same results were obtained for many sequences of input pulses, in different amplitudes and frequencies. To quantify this smoothing feature, we compared the response of the bifan motif to different input signals. We stimulated the circuit by an oscillatory signal, such as $\sin \omega t$, and calculated the extent to which the output is oscillating with the same frequency ω. It is expected that a filter will reduce the presence of the input frequency in the output. The mathematical way of measuring the presence of a frequency ω in a general function (of time) is by calculating its Fourier transform. Thus we would say that a circuit A is a better filter than circuit B, if the ω Fourier component in the output of A is smaller than that of B. We show an example in Fig. 4b, where we compare two circuits – the bifan motif with AND gates, and a similar circuit with OR gates. The ATF2 homodimer production rate is the output of the circuits. For each circuit we present the power spectrum, namely the square of Fourier transform of the output resulted from stimulation by a constant input and by an oscillatory input with period of 10 minutes. The presence of the frequency ω in the output is seen as a sharp peak in its Fourier transform. This peak appears only at the response to oscillatory input, but not in the output obtained by a constant input. It is evident that the input frequency is significantly present in the output from OR gated bifan, much more than in the AND bifan output. We compare the output resulted by an oscillatory input to that of a constant input. In the OR gate circuit the component which relates to the input frequency was enhanced by more than 1000 fold. The AND bifan circuit filtered out a significant part of the oscillations and the enhancement is less than 50-fold. Same analysis was applied for a wide range of oscillations (with periods from 30 seconds to more than 1.5 hours). For each oscillatory input with frequency ω, we find the ω Fourier component of the output, and calculate its square $F_{osc}(\omega)$. We compare this value to $F_{const}(\omega)$ - the square of ω Fourier component of the output after stimulation by a constant signal. We define the circuit's filtering index for frequency ω as the ratio $F_{osc}(\omega)/F_{const}(\omega)$. This ratio quantifies the extent to which the circuit enables the incoming oscillations to be transferred and reflected by the output. The normalization by $F_{const}(\omega)$ makes sure that we calculate only the circuit "transparency" to oscillation, and not affected by the overall performance of the circuit. High value indicates poor filtering, and an ideal filter, which smoothes the output completely, has a filtering index of 1. Comparison between the filtering index of several circuits indicates that the AND gate configuration gives the best (i.e. smallest) filtering index for almost any signal frequency, and in many cases the difference is of several orders of magnitude (Fig. 4c).

To understand the effects of having a bifan motif with the additional arm from JNK to c-Jun, we compared the dynamics of the original configuration (Fig. 1b) to a system with the same topology except that we eliminated the cross links forming the bifan. In one set of simulations, we removed the links between p38α to ATF2 and the link between JNK to Elk-1, and in another set of simulation we removed the links between p38α and Elk-1 and between JNK and ATF2 (Fig. 5). When the cross links are eliminated, each transcription factor is regulated independently by one protein kinase. In those altered configurations there is no coordination between the activation routes and thus the two transcription factors can display different dynamics (Fig. 5, two bottom panels). Interestingly, only when all links are present the activation of Elk-1 and ATF2 is synchronized and the two transcription factors exhibit similar concentrations at any time point.

To determine if synchronization is a robust property, we performed several sets of simulations, with a broad range of initial conditions. We varied the ratio between the initial condition of p38 to that of JNK from 10:1 to 1:10 and measured the synchronization between the time course of active transcription factors. The synchronization was calculated as follows: Time course of ATF2* and ELK* was normalized such that each of the TFs would have maximal value of 1. The square of the difference between the two TF concentrations was averaged over time. The square root of this average is a measure to the deviation from synchronization. Perfect synchronization yields no difference between the (normalized) TF concentrations, and thus the deviation is zero. Uncorrelated TFs would yield high value of deviation. In Fig. 6 we show the deviation from synchronization for various initial conditions, in the three configurations: full bifan, and two options of removing the cross links. For the whole range of parameters, the bifan configuration exhibit better synchronization (smaller deviation) than the other two configurations. The synchronization feature of the bifan configuration is more significant under conditions of low p38:JNK ratio. This feature is not dependent on exact tuning of the parameters. To verify this, we changed the reaction rates of p38. We examined range of values between 0.75- to 1.35-fold of the original rates. We varied the complex formation rates ($k_3$ and $k_9$ in Table II), the complex dissociation rates ($k_4$, $k_5$, $k_{10}$, and $k_{11}$) or both. In all cases we obtained similar results. In addition, we varied the ratio between the concentrations of the protein kinases and the transcription factors. The results are shown in Fig. 7. As before, the fully linked bifan synchronizes the TF activity better than the other configurations. Furthermore, for the fully linked bifan the synchronization quality is largely independent on the TF/kinase ratio, as compared to the other configurations. These results indicate that synchronization is a robust property of bifans.

We next analyzed a more complex network that contained the bifan motif and the two known feedback loops involving IEG: The positive feedback loops of c-Jun to itself (51, 61) and the negative feedback loop involving MKP-1 (53-55). First, we "added" the positive feedback loop involving c-Jun. Either ATF2:c-Jun heterodimers or c-Jun homodimers can initiate c-Jun transcription. Other proteins in the circuit were assumed to be in steady state, so their transcription, translation and degradation were not explicitly simulated. Production and degradation were considered for c-Jun only. It is known that phosphorylation of c-Jun results in protection of c-Jun from ubiquitination and ubiqutin-dependent degradation (62). Thus, degradation was considered for the non-active form only. The rates for the simulations were chosen to maintain promoter occupancy that would have a significant effect on c-Jun production, while being sensitive to changes in input signal (Table 3). Thus, we allowed sufficient promoter activation to influence the dynamics of the entire network. The parameters for the bifan motif were the same as in previous simulations (see Methods and Table 2). As before, this circuit was simulated with and without the cross links (Fig.8). Unexpectedly, the temporal profile of activation for Elk-1 and ATF2 are similar to what is observed for the smaller circuit (Fig.5). However due to the presence of the positive feedback loop, c-Jun activation is prolonged. The fact that the ATF2 activation is not synchronized with c-Jun causes excess production of homo-dimers in respect to ATF2:c-Jun hetrodimers. These results suggest that the target activity synchronization is an intrinsic feature of the bifan, whereas the c-Jun positive feedback is an independent module, separated from the core bifan.

The negative feedback loop involving the IEG phosphatase MKP-1 was "added" to the circuit (Fig. 8). Transcription of MKP-1 is regulated by ATF2:c-Jun hetrodimers and ATF2 homodimers (Fig. 1d). The results of the simulations of this circuit show that the temporal profiles are significantly affected. There is a sharp increase in activity followed by a relatively rapid decline due to the negative feedback loop. Nevertheless the effect of removing the cross links remains similar to that of the less complex circuits (Fig. 9). As in previous results (Figs. 5 and 8), the synchronization feature of the bifan is maintained, regardless of the surrounding network. Removal of the cross links, on the other hand, leads to asynchronized dynamics. We conclude that synchronization, i.e. correlated activation of transcription factors is a fundamental feature intrinsic to the bifan motif.

We next explored the relationship between the bifan and the negative feedback loop. The bifan motif is nested within the MKP-1 negative feedback loop (Fig 10a). Replacing the bifan with a single node yields a standard feedback loop. Two of the bifan products (ATF2 homodimer and ATF2:c-Jun hetrodimer) are required for initiating MKP-1 transcription, which leads to deactivation of the protein kinase p38 and JNK. What would be the effect of the bifan on the negative feedback loop? One could assume that high initial concentration of MKP-1 will repress the bifan activity. The results which are presented in Fig. 11 show that this is not the case. When the bifan motif is fully connected, it takes some time to inhibit the bifan's function, and thus, MKP-1 can be produced (Fig. 11a). On the other hand, when the cross links of the bifan are deleted, the initial amount of MKP1 can alter this dynamics, the output products do not initiate the MKP-1 synthesis, making the feedback loop non-operational (Fig. 11b). Different initial conditions may be a result of system's history. Under conditions of dynamic environment with many incoming signals, the cell can be found in various states. Our results indicate that the bifan motif may serve as a control unit, enabling proper operation of other motifs under different conditions.

## DISCUSSION

*Upstream regulation of the JNK/p38 protein kinase bifan*

It should be noted that although, p38 and JNK are regulated by complex upstream signaling network. For example, it is known that Cdc42 is a key GTPase regulating the activation of both JNK and p38 (63). JNK can be activated by large transient Ca++ waves (64), whereas JNK1 and JNK2 are known to be activated by MEKK1 (65). Additionally, the MAPK1/2 pathway sometimes functions antagonistically to the JNK and p38 pathway. Thus, when one pathway is active, sometimes the other is inactive and vise-versa. These pathways in certain cellular scenarios may have opposite effects on cellular phenotypes (66). The bifan motif can play a crucial role in transmitting this rich complexity of upstream signaling to downstream targets. When we extended the bifan motif to include the IEG c-Jun and MKP-1, and compared the original configuration to alternative topologies that do not implement a bifan motif, but contain enough necessary links for functional connectivity, we observed markedly different concentrations of homodimers and hetrodimers. Hence, bifan motifs may be necessary to provide balance and signal sorting to ensure the production of proper levels of activated transcription factor combinations of homodimer and hetrodimers for the appropriate regulation of transcription. We have found that under certain conditions bifan motifs can filter noise and synchronize the activity of activated transcription factors. Additionally, we show that the signal processing capacity of bifan motifs is highly dependent on the context in which they function, where the initial concentrations of different circuit components can markedly affect the dynamical behavior. The quantitative analysis of cell-signaling circuits, coupled to gene-regulation which feeds back into the cell-signaling network is in itself an important advancement towards more complex modeling of the dynamics of intracellular regulation.

*Functional effects of multiple isoforms*
The abundance of bifan motifs in intracellular regulatory networks is mostly due to the presence of isoforms, protein with similar sequence and function, found abundantly in mammalian cells (67). A common configuration of bifan motifs consists of two upstream regulators that are isoforms and two substrates that are isoforms. Isoforms are thought to be created through an evolutionary process of duplication and divergence, and artificial growing network models, that use the duplication and divergence for network growth (68), were shown to produce networks with characteristics similar to those of cell signaling networks (69) .

*AND vs. OR configuration of bifans*
Some of the advantages listed above depend on the AND/OR configuration. For example, the reliability mechanism is relevant only for the OR configuration whereas AND configurations provides the opposite advantage: coincidence detection. AND gating makes sure that the substrates would not be activated accidentally by random signals. Similar conclusions were arrived by Mangan and Alon (8) analyzing OR and AND configurations for feed forward motifs, suggesting that motif dynamics are mostly influenced by the logic gating impinging on source nodes. Thus, advantages related to information processing are coupled to topological configuration details that go beyond just links and nodes.

*Evidence for AND gating*
Although the gating mechanisms of activation for the transcription factors studied, ATF2, Elk-1 and c-Jun are unclear experimentally, there are several examples of sequential phosphorylations of transcription factors by multiple protein kinases supporting the ordered AND gating possibility. For example, MAPK1/2 initial phosphorylation of serine 307 of heat shock factor-1 (HSF-1) is required for the later suppression of activity by the sequential phosphorylation of serine 303 by the protein kinase glycogen synthase kinase 3 (GSK3) (70). Another example is the phosphorylation of tau by protein kinase A (PKA) and GSK3β. In order to be recognized by the antibody AT100, the tau protein must be phosphorylated first by GSK3β at Thr212 and then by PKA at Ser214. If Ser214 is phosphorylated first, it protects Thr212 from being phosphorylated (71). Using a synthetic peptide it has been shown that GSK3 consensus phosphorylation amino-acid-substrate-target-sequence requires prior phosphorylation by casein kinase II (72). Ruzzene et al. (73) showed that Syk phosphorylation of HS1 potentiates this protein to be a good c-Fgr (and other Src tyrosine protein kinase family members) substrate. The retinoblastoma protein (pRb) is an important inhibitor of the cell cycle. For cells to enter G1 from S phase, Rb is inhibited by double sequential phosphorylations, first by cyclin

D-cdk4/6 complexes and then by cyclin E-cdk2 complexes (74). It is also known that c-Jun is phosphorylated on both serine 63 and serine 73 by JNK (44-46). Whether these phosphorylations are sequential is not yet clear, but the above examples suggest that ordered AND gating is likely to be a common mechanism used for signal information processing.

*Placing motifs in context: coupling of motifs within networks*
Quantitative analysis of network motifs should include the effects of placing motifs in context of other motifs. A convenient way of distinguishing between local and global effects is by gradual addition of links and nodes while comparing the outputs obtained from different levels while expanding the network. In our case, analysis of the bifan motif without considering the context in which it is embedded can miss important aspects of the motif's quantitative dynamical behavior. As suggested by Ingram et. al. (22), the bifan motif can display a range of behaviors, not encoded within the abstraction to nodes and links alone. In this study, we extended the analysis to include additional components and interactions and placed the bifan in context of larger networks. This allowed us to identify the unique contribution of the bifan motif to the network functional performance and to find new emergent properties of the bifan motif such as synchronization and filtering. These properties were previously reported for other motifs and as such place the bifan in functional context of other regulatory motifs (Table 1).

How different motifs are juxtaposed next to each other and the consequences of coupling between motifs is the next step in understanding the structure-function relation in networks. Motifs, as network elements, can be coupled in many different ways, stacked in serial or parallel combinations, or nested within one another. A bifan with nested feed forward loops that emerge from the two output nodes was analyzed in the context of signaling networks and was named a multi-layered perceptron (15). Our extended network (Fig. 1d) contains two examples of coupled motifs. The c-Jun positive feedback loop is serially stacked with the bifan since the output of the bifan motif is the input for the positive feedback loop (Fig 10b). In this configuration the synchronization behavior of the bifan and the extended activation of c-Jun by the positive feedback loop are observed. In contrast, the bifan motif is nested within the MKP1 negative feedback loop (Fig 10a). In this configuration, when the bifan motif is fully connected, significant amount of MKP1 protein is synthesized, making the negative feedback loop operational. When the cross links of the bifan are deleted, the output products do not initiate the MKP1 synthesis, making the feedback loop non-operational. Thus, in this simple model of two interacting motifs, the presence of the one motif is essential for the existence and function of the second motif. Regulatory network motifs are similar to resistors and capacitors in electrical circuits (6). Understanding the basic laws of each individual element is a fundamental prerequisite for quantitative analysis of large and complex signaling and transcription regulatory circuits. Full system understanding needs to include both intrinsic quantitative properties of motifs and how interactions between motifs lead to reciprocal effects.

**MATERIALS AND METHODS**

*Rate constants*

Rate constants were obtained or estimated from experimental studies that describe the kinetics of various *in-vitro* phosphorylation reactions. These rates are provided in Table 2. Kinetic rates for the phosphorylation of ATF2 by p38α were taken from LoGrasso et al. (75). The phosphorylation rate constant of Elk-1 by p38α was based on comparative rates derived from Goedert et al. (76). The kinetic rate constant for JNK1 phosphorylation of c-Jun was derived using Lineweaver-Burke double reciprocal plots from the data presented by Kallunki et al. (49). Reaction rates of JNK1 phosphorylation of ATF2 and Elk-1 were assumed to be the same based on Gupta et al. (48). Time course data for c-Jun (77, 78) and ATF2 activation (78-80) suggests that the transcription factors are active for about 120 minutes on average, reaching their peak at between 15 and 30 minutes. Unknown rates were adjusted to follow this dynamics. $K_{on}$ and $K_{off}$ binding constants were set to favor dimer formation. The rates were set so the promoters will be occupied during some fraction of time, making the feedback loop effective, but not saturated.

A summary of constants which were used for the simulations presented in the Figures are provided in Table 3. In addition to the parameters listed in Table 3, range of parameters were applied to make sure that the results are insensitive to exact values.

*Mathematical model for protein kinase – transcription factor interactions*

A set of ordinary differential equations (ODEs) was written and numerically solved to follow the time course of concentrations. The equations follow the dynamics of all components of the network, as well as intermediate complexes. AND and OR gates were modeled by having different protein forms and intermediate complexes, respectively. For example, in the OR configuration each TF had one equation for the unphosphorylated form, one equation for the active phosphorylated form, and two more equations for the two intermediate complexes (one with each of the kinase proteins). In the AND configuration, there are three equations for the phosphorylated form – one for a TF phosphorylated by JNK, one for a TF phosphorylated by p38α and a third one for the doubly phosphorylated TF, and the list of intermediate complexes was set accordingly. The differences between the species and reactions involved in any of the configurations are summarized in Table 2. In the ordered AND version, activation by p38α preceded activation by JNK.

*Model for transcription*

For the regulated transcription in the feedback loops, the dynamics of the occupied promoters was simulated explicitly, rather than using Michaelis-Menten approximation. The Michaelis-Menten equations are based on the assumption that reaction rates are fast enough to treat biomolecular components as concentrations. Furthermore, intermediate complexes are often assumed to be in steady state. These assumptions do not hold for active promoters since there are only two copies, and at such low copy numbers discretization and fluctuations have a significant effect on the dynamics. Thus, we did not assume steady state for the promoter occupancy. Instead, we extended the set of ODEs to include the promoter as a separate species. Its occupancy is time dependent, and is governed by an equation consists of binding and unbinding terms (81). The actual transcription rate is assumed to be proportional to the occupancy of the promoter. The limited number of promoters (one or two per cell) was modeled by defining a maximal concentration of occupied promoters, which is the number of promoters converted into units of concentration. The binding rate is proportional to the concentration of unoccupied promoters, namely the difference between maximal and actual occupancy. This way the concentration of occupied promoters cannot exceed its upper limit. For simplicity, transcription and translation were modeled as a single step.

*Regulation of transcription by multiple promoters*
In the MKP1 loop, the two promoters have to be occupied simultaneously (AND gate) to initiate MKP1 transcription. This was modeled by calculating the MKP1 actual transcription rate as the maximal transcription rate multiplied by the concentrations of each the occupied promoters. Thus, it is enough to have one promoter unoccupied (so that the concentration of the respective occupied promoter is zero) in order to prevent the transcription from occurring.
OR gate was assumed to regulate the c-Jun transcription. Namely, gene transcription becomes possible by occupying any of the regulatory elements. In that case, the transcription term should be proportional to the sum of the promoters' occupancy, reflecting the fact that one occupied promoter is enough.

All simulations were performed using the standard ODE solver of Matlab$^{TM}$ (Natick, MA). All scripts m files are provided as online supplementary materials and on the Iyengar Laboratory web-site at http://www.mssm.edu/labs/iyengar/resources.

## ACKNOWLEDGEMENTS

This research is supported by NIH Grants GM-054508 and GM-072853 and an Advanced Center Grant from NYSTAR.
The authors have declared that no competing interests exist.

| Functional Capabilities | **Bifan** (This study) | **Feed Forward Loop** | **Feedback Loop** | **SIM** |
|---|---|---|---|---|
| **Delay/Speedup** | Depends on logical gating | Exact function depends on sign, logical gating, and coherence (8, 13, 16, 24) | Autoregulation may reduce response time whereas longer loops can cause delay. (14, 33) | |
| **Filter noise** | AND gates filter better than OR gates. | FFLs reject transient input pulses and respond only to persistent stimuli (17, 25) | Filtering by using interlinked loops (29) | |
| **Sort and Synchronization** | TF activity is synchronized under a broad range of conditions | | | The input node is often autoregulated, yielding a coupling between a FBL and a SIM (13, 14) |
| **Bistability** | | | Fluctuations and stochasticity play a role in switching between the stable states (18, 20, 21, 26-28). | |
| **Oscillations** | | | Negative feedback is required for oscillations(12, 18, 19, 29-32). | |

**Table 1. Comparison of functional capabilities demonstrated quantitatively for different regulatory network motifs.**

| Reaction | Rate (expression) | OR | AND | Ordered AND | Rate (value) | Reference |
|---|---|---|---|---|---|---|
| Signal1 + p38 → p38* | [signal1][p38] | X | X | X | included in signal | |
| p38* → p38 | $k_1$[p38*] | X | X | X | $k_1$=0.1 | |
| Signal2 + JNK → JNK* | [signal2][JNK] | X | X | X | included in signal | |
| JNK* → JNK | $k_2$[JNK*] | X | X | X | $k_2$=0.1 | |
| p38* + ATF2 → p38*:ATF2 | $k_3$ [p38*][ATF2] | X | X | X | $k_3$=3.9 | LoGrasso PV, Frantz B, Rolando AM, O'Keefe SJ, Hermes JD, O'Neill EA. Kinetic mechanism for p38 MAP kinase. Biochemistry. 1997 36:10422-7 |
| p38* + ATF2$^{JNK}$ → p38*:ATF2$^{JNK}$ | $k_3$[p38*][ATF2$^{JNK}$] | | X | | | |
| p38*:ATF2 → p38* + ATF2 | $k_4$ [p38*:ATF2] | X | X | X | $k_4$=19.2 | |
| p38*:ATF2$^{JNK}$ → p38* + ATF2$^{JNK}$ | $k_4$ [p38*:ATF2$^{JNK}$] | | X | | | |
| p38*:ATF2 → p38* + ATF2* | $k_5$ [p38*:ATF2] | X | | | $k_5$=4.8 | |
| p38*:ATF2 → p38* + ATF2$^{p38}$ | $k_5$ [p38*:ATF2] | | X | X | | |
| p38*:ATF2$^{JNK}$ → p38* + ATF2* | $k_5$ [p38*:ATF2$^{JNK}$] | | X | | | |
| JNK* + ATF2 → JNK*ATF2 | $k_6$ [JNK*][ATF2] | X | X | | $k_6$=20 | Gupta S, Barrett T, Whitmarsh AJ, Cavanagh J, Sluss HK, Derijard B, Davis RJ. Selective interaction of JNK protein kinase isoforms with transcription factors. EMBO J. 1996 15:2760-70 |
| JNK* + ATF2$^{p38}$ → JNK*:ATF2$^{p38}$ | $k_6$ [JNK*][ATF2$^{p38}$] | | X | X | | |
| JNK*:ATF2 → JNK* + ATF2 | $k_7$ [JNK*:ATF2] | X | X | | $k_7$=40 | |
| JNK*:ATF2$^{p38}$ → JNK* + ATF2$^{p38}$ | $k_7$ [JNK*:ATF2$^{p38}$] | | X | X | | |
| JNK*:ATF2 → JNK* + ATF2* | $k_8$ [JNK*:ATF2] | X | | | $k_8$=10 | |
| JNK*:ATF2 → JNK* + ATF2$^{JNK}$ | $k_8$ [JNK*:ATF2] | | X | | | |

| Reaction | Rate | | | | Rate constant | Reference |
|---|---|---|---|---|---|---|
| JNK*:ATF2$^{p38}$ → JNK* + ATF2* | $k_8$ [JNK*:ATF2$^{p38}$] | | X | X | | (Assumed to be the same as JNK-c-Jun rates) |
| p38* + ELK-1 → p38*:ELK-1 | $k_9$ [p38*][ELK-1] | X | X | | | Goedert M, Cuenda A, Craxton M, Jakes R, Cohen P. Activation of the novel stress-activated protein kinase SAPK4 by cytokines and cellular stresses is mediated by SKK3 (MKK6); comparison of its substrate specificity with that of other SAP kinases. EMBO J. 1997 16:3563-71 |
| p38* + ELK-1$^{JNK}$ → p38*:ELK-1$^{JNK}$ | $k_9$[p38*][ELK-1$^{JNK}$] | | X | X | $k_9$=16.02 | |
| p38*:ELK-1 → p38* + ELK-1 | $k_{10}$ [p38*:ELK-1] | X | X | | | |
| p38*:ELK-1$^{JNK}$ → p38* + ELK-1$^{JNK}$ | $k_{10}$ [p38*:ELK-1$^{JNK}$] | | X | X | $k_{10}$=35 | |
| p38*:ELK-1 → p38* + ELK-1* | $k_{11}$ [p38*:ELK-1] | X | | | | |
| p38*:ELK-1 → p38* + ELK-1$^{p38}$ | $k_{11}$ [p38*:ELK-1] | | X | | | |
| p38*:ELK-1$^{JNK}$ → p38* + ELK-1* | $k_{11}$ [p38*:ELK-1$^{JNK}$] | | X | X | $k_{11}$=8.75 | |
| JNK* + ELK-1 → JNK*:ELK-1 | $k_{12}$ [JNK*][ELK-1] | X | X | X | | Gupta S, Barrett T, Whitmarsh AJ, Cavanagh J, Sluss HK, Derijard B, Davis RJ. Selective interaction of JNK protein kinase isoforms with transcription factors. EMBO J. |
| JNK* + ELK-1$^{p38}$ → JNK*:ELK-1$^{p38}$ | $k_{12}$ [JNK*][ELK-1$^{p38}$] | | X | | $k_{12}$=20 | |
| JNK*:ELK-1 → JNK* + ELK-1 | $k_{13}$ [JNK*:ELK-1] | X | X | X | | |
| JNK*:ELK-1$^{p38}$ → JNK* + ELK-1$^{p38}$ | $k_{13}$ [JNK*:ELK-1$^{p38}$] | | X | | $k_{13}$=40 | |
| JNK*:ELK-1 → JNK* + | $k_{14}$ [JNK*:ELK-1] | X | | | $k_{14}$=10 | |
| ELK-1* | | | | | | 1996 15:2760-70 |
| JNK*:ELK-1 → JNK* + ELK-1$^{JNK}$ | $k_{14}$ [JNK*:ELK-1] | | X | X | | (Assumed to be the same as JNK-c-Jun rates) |
| JNK*:ELK-1$^{p38}$ → JNK* + ELK-1* | $k_{14}$ [JNK*:ELK-1$^{p38}$] | | X | | | |
| ATF2* + ATF2* → ATF2*:ATF2* | $k_{15}$ [ATF2*]$^2$ | X | X | X | $k_{15}$=0.02 | |
| ELK-1* + ELK-1* → ELK-1*:ELK-1* | $k_{16}$ [ELK-1*]$^2$ | X | X | X | $k_{16}$=0.02 | |

**Table 2. Reactions and rates used to simulate the bifan motif. Some of the reactions occur at all variants of the motif whereas others are used only at one or two configurations. X$^Y$ represents a substrate X which is phosphorylated by kinase Y. This substrate is not considered activated until a second phosphorylation by another kinase. Full activation is denoted by X*. This may be either activation by a single kinase in the OR gate configuration, or double phosphorylation in the AND gate cases. All rates for the reactions of the form A+B→… are in $\mu M^{-1} min^{-1}$. Rates for the A→… reactions are in $min^{-1}$.**

|  | **Reaction** | **Rate (expression)** | **Rate (value)** |
|---|---|---|---|
| Additional leg (Fig. 1b) | JNK* + c-Jun → JNK*:c-Jun | $k_{17}$ [JNK*][c-Jun] | $k_{17}$=20 |
|  | JNK*: c-Jun → JNK* + c-Jun | $k_{18}$ [JNK*: c-Jun] | $k_{18}$=40 |
|  | JNK*: c-Jun → JNK* + c-Jun* | $k_{19}$ [JNK*: c-Jun] | $k_{19}$=10 |
|  | c-Jun * + ATF2* → c-Jun:ATF2 | $k_{20}$ [ATF2*][c-Jun*] | $k_{20}$=0.02 |
|  | c-Jun * + c-Jun * → c-Jun:c-Jun | $k_{21}$ [c-Jun *]$^2$ | $k_{21}$=0.02 |
| Negative feedback loop | ATF2:ATF2 + MKP-1_P1 → MKP1_P1:ATF2$^2$ | $k_{22}$ [ATF2:ATF2]($P_{max}$-[MKP1_P1:ATF2$^2$]) | $k_{22}$=5 |
|  | c-Jun:ATF2 + MKP-1_P2 → MKP1_P2:c-Jun:ATF2 | $k_{22}$ [c-Jun:ATF2]($P_{max}$-[MKP1_P2:c-Jun:ATF2]) |  |
|  | MKP1_P1: ATF2$^2$ → ATF2:ATF2 + MKP-1_P1 | $k_{23}$[MKP1_P1:ATF2$^2$] | $k_{23}$=1 |
|  | MKP1_P2:c-Jun:ATF2 → c-Jun:ATF2 + MKP-1_P2 | $k_{23}$[MKP1_P1:c-Jun:ATF2] |  |
|  | ∅ → MKP-1 | $k_{24}$[MKP1_P1:ATF2$^2$][MKP1_P2:c-Jun:ATF2] | $k_{24}$=3 |
|  | MKP-1 → ∅ | $k_{25}$[MKP-1] | $k_{25}$=0.05 |
|  | MKP-1 + p38* → MKP-1 + p38 | $k_{26}$[MKP-1][p38*] | $k_{26}$=0.1 |
|  | MKP-1 + JNK1* → MKP-1 + JNK1 | $k_{27}$[MKP-1][JNK1*] | $k_{27}$=0.1 |
| Positive feedback loop | c-Jun:ATF2 + c-Jun_P1 → c-Jun_P1:c-Jun:ATF2 | $k_{28}$ [c-Jun:ATF2]($P_{max}$-[c-Jun_P1:c-Jun:ATF2]) | $k_{28}$=20 |
|  | c-Jun:c-Jun+ c-Jun_P2 → c-Jun_P2:c-Jun$^2$ | $k_{28}$ [c-Jun:c-Jun]($P_{max}$-[c-Jun_P2:c-Jun$^2$]) |  |
|  | c-Jun_P1:c-Jun:ATF2 → c-Jun:ATF2 + c-Jun_P1 | $k_{29}$ [c-Jun_P1:c-Jun:ATF2] | $k_{29}$=10 |
|  | c-Jun_P2:c-Jun$^2$ → c-Jun$^2$ + c-Jun_P2 | $k_{29}$ [c-Jun_P2:c-Jun$^2$] |  |
|  | ∅ → c-Jun | $k_{30}$($k_{31}$+[c-Jun_P2:c-Jun$^2$]/$P_{max}$+[c-Jun_P1:c-Jun:ATF2]/$P_{max}$) | $k_{30}$=1 $k_{31}$=1 |
|  | c-Jun → ∅ | $k_{32}$[c-Jun] | $k_{32}$=0.1 |
| Dimer degradation | ATF2:ATF2 → ∅ | $k_{33}$ [ATF2:ATF2] | $k_{33}$=0.2 |
|  | c-Jun:ATF2 → ∅ | $k_{34}$ [c-Jun:ATF2] | $k_{34}$=0.2 |
|  | c-Jun:c-Jun → ∅ | $k_{35}$ [c-Jun:c-Jun] | $k_{35}$=0.2 |

**Table 3. Reactions and rates used for the extended circuit of Figs. 1c and 1d. The number of promoters was modeled by limiting the concentration of the bound promoters to an upper limit of $P_{max}$=0.001 $\mu M^{-1}$. All rates for**

reactions of the form A+B→… are in $\mu M^{-1} min^{-1}$. Rates for the A→… reactions are in $min^{-1}$.

**FIGURE LEGENDS**

**Fig. 1**

The JNK1/p38 – ATF2/Elk-1 bifan motif configuration and its immediate environment. a) The basic motif is in the shaded area. b) The shaded area is the bifan motifs with an adjacent link from JNK to c-Jun. JNK phosphorylates and activates c-Jun. c) The shaded area includes a positive feedback loop involving the immediate early gene c-Jun. d) The shaded area includes a negative feedback loop involving the immediate early gene MKP-1.

**Fig. 2**

(a) Truth table and diagram of the OR gate. (b) Truth table and diagram of the AND gate. (c) Concentration of free active ATF2 as a function of time for various configurations of the bifan motif shown in Fig. 1a. Stimulus was given at time 0<t<5. The initial condition was: [p38] = [JNK] = 10 μM, [ATF2] = [ELK1] = 30 μM. (d) Concentration of free active ATF2 as a function of time for various configurations of the bifan motif shown in Fig. 1a. Stimulus was given for one minute every 5 minutes (periodically). The initial conditions were the same as in (c). (e) Concentration of free active ATF after a single 1 minute pulse.

**Fig. 3**

Effect of increasing the initial concentration of Elk-1 on ATF2 homodimers production. Elk-1 "trapping" of JNK1 and p38α, leaving less free p38* (a), results in a decrease in the production of activated ATF2 homodimers (b). Non ordered AND gate was assumed, Initial condition: [p38] = [JNK] = 10 μM, [ATF2] = 30 μM. Initial concentration of Elk-1 varies between 0 and 50μM, as indicated by the curves. The Figure presents the concentration of produced homodimers disregarding reactions such as degradation, dissociation etc..

**Fig. 4**

The bifan motif as a filter for fluctuating signals. Time course of ATF2 homodimers is the output. (a) An example of random pulse series as input. (b) Filtering periodic signal. The power spectrum of the output from oscillating signal to the OR gate motif (thick dashed line, right panel) deviates from the response to the non-oscillating input (thin dashed). The deviation at the AND gate configuration (solid lines, left panel) is much smaller. (c) Oscillation filtering index $F_{osc}(\omega)/F_{const}(\omega)$ (see text for definitions) of various configurations as a function of time period.

**Fig. 5**

Dynamics of activated transcription factors in a bifan motif configuration with an additional arm from JNK to c-Jun. The original configuration as shown in Fig. 1b (top panel); after eliminating the links from p38 to ELK1 and from JNK1 to ATF2 (middle panel); and without the links from p38 to ATF2 and from JNK1 to ELK1 (bottom panel). The initial conditions were: [p38] = [JNK] = 3 μM, [ATF2] = [ELK1] = 30 μM, and [c-Jun] = 10 μM. OR gates were assumed for all substrate activations. Cartoons of the network are presented by each panel. The signal inputs are presented at the most upper panel.

**Fig. 6**

Deviation from synchronization of bifan motif and non-bifan circuits for a range of initial conditions. Deviation from synchronization was calculated as the root mean square of the difference between normalized concentrations (see text for details). The network is the same as in Fig. 5 and the initial conditions are: [ATF2]=[ELK]=20μM, [c-Jun]=10μM, [p38]=3μM, and [JNK] was varying from 0.3 to 30 μM. As an example, time course of free ATF2* and ELK* is presented in the inset for the case `case p38:JNK = 1:10`. In the full bifan configuration (solid lines) the two lines coincide, whereas without the cross links (dotted line) the two TFs are not synchronized.

**Fig. 7**

Deviation from synchronization of bifan motif and non-bifan circuits for a range of initial conditions. The network is the same as in Fig. 5. The initial conditions are: [p38]= [JNK] =3μM, [ATF2],[ELK], and [c-Jun] varying between 10 μM to 50 μM, yielding TF/kinase ratios between 3 and 17.

**Fig. 8**

Dynamics of free monomeric activated transcription factors (TF) in a bifan motif configuration adjacent to a positive feedback loop. The full bifan configuration as shown in Fig. 1c (top panel); after eliminating the links from p38 to ELK1 and from JNK1 to ATF2 (middle panel); and without the links from p38 to ATF2 and from JNK1 to ELK1 (bottom panel). The initial conditions were: [p38] = [JNK] = 3 μM, [ATF2] = [ELK1] = 30 μM, and [c-Jun] = 10 μM. OR gates were assumed for all substrate activations. Cartoons of the network are presented by each panel. The signal inputs are presented at the most upper panel.

**Fig. 9**

Dynamics of free monomeric activated transcription factors (TF) in a bifan motif configuration adjacent to a positive feedback loop and nested in a negative feedback loop. The full bifan configuration as shown in Fig. 1d (top panel); after eliminating the links from p38 to ELK1 and from JNK1 to ATF2 (middle panel); and without the links from p38 to ATF2 and from JNK1 to ELK1 (bottom panel). The initial conditions were: [p38] = [JNK] = 3 $\mu$M, [ATF2] = [ELK1] = 30 $\mu$M, and [c-Jun] = 10 $\mu$M. OR gates were assumed for all substrate activations. Cartoons of the network are presented by each panel. The signal inputs are presented at the most upper panel.

**Fig. 10**

Coupling of two motifs. (A) Nesting of a bifan motif in a negative feedback loop. (B) Serial combination – the output of the bifan is the input for the positive feedback loop

**Fig. 11**

Effect of initial condition of MKP1 levels on the feedback loop dynamics. When the bifan synchronizes the dynamics of the transcription factors, the feedback loop is highly active (upper panel). However, after eliminating the links from p38 to ELK1 and from JNK1 to ATF2 there is no more synchronization, and thus no transcription of MKP1 observed (lower panel).

Fig. 1

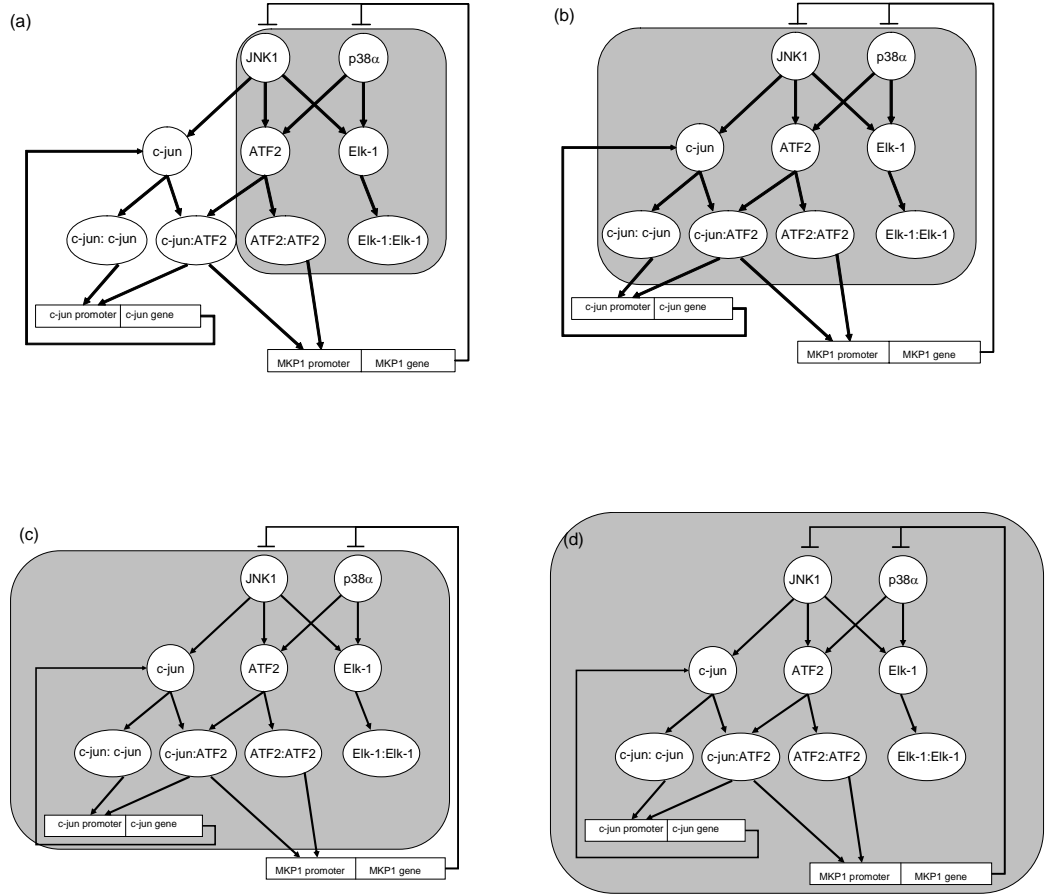

**Figure 2**

(a)

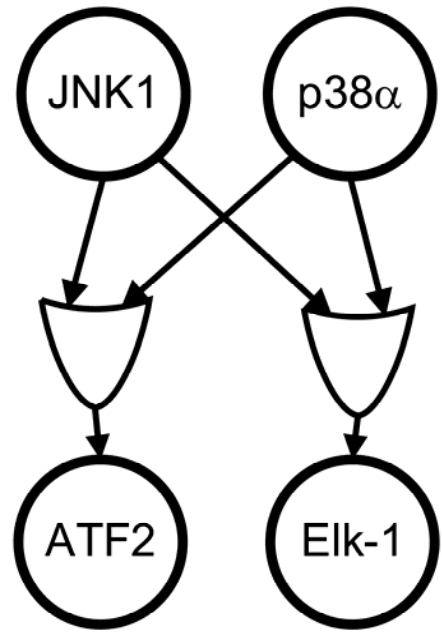

| JNK \ p38 | Inactive | Active |
|---|---|---|
| Inactive | 0 | 1 |
| Active | 1 | 1 |

(b)

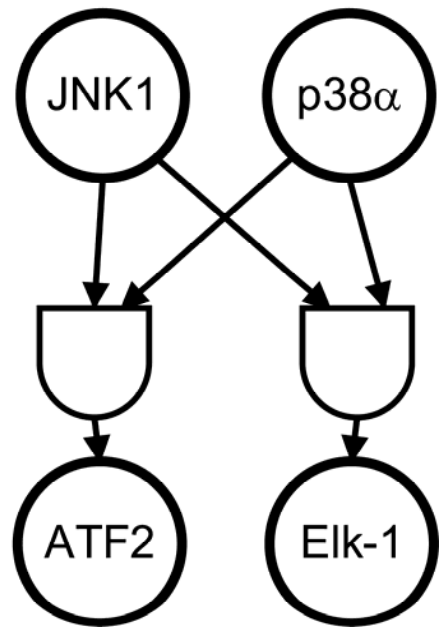

| JNK \ p38 | Inactive | Active |
|---|---|---|
| Inactive | 0 | 0 |
| Active | 0 | 1 |

**Fig. 2c**

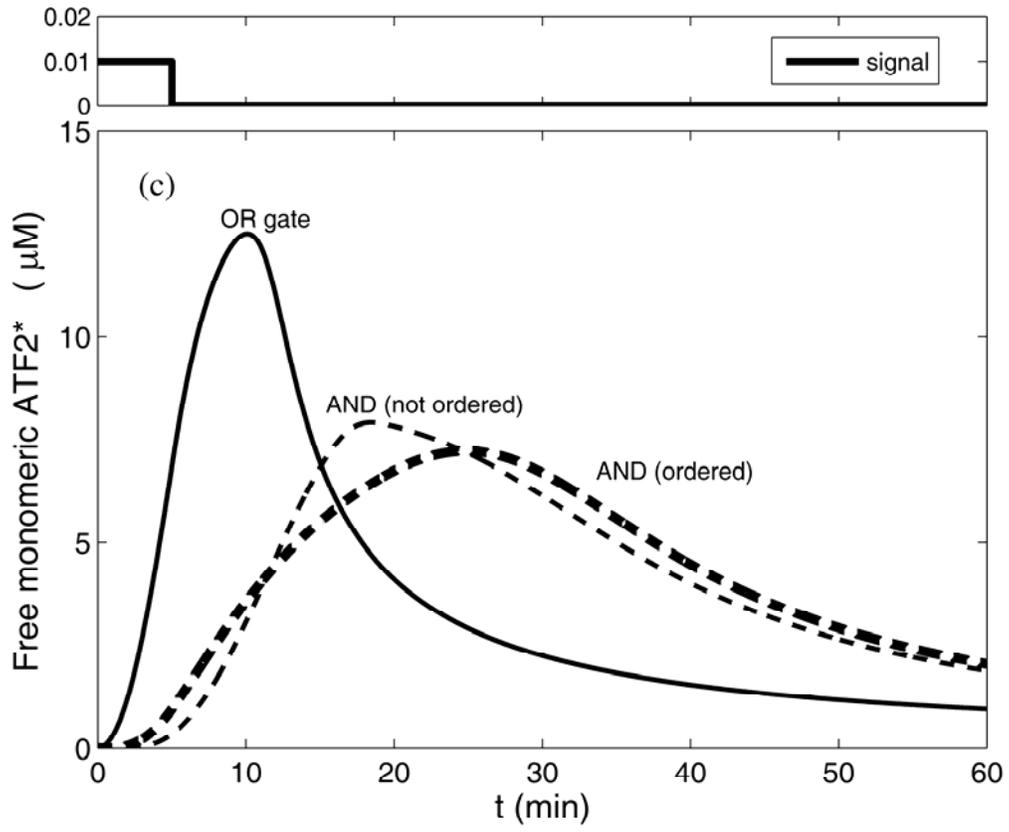

**Fig. 2d**

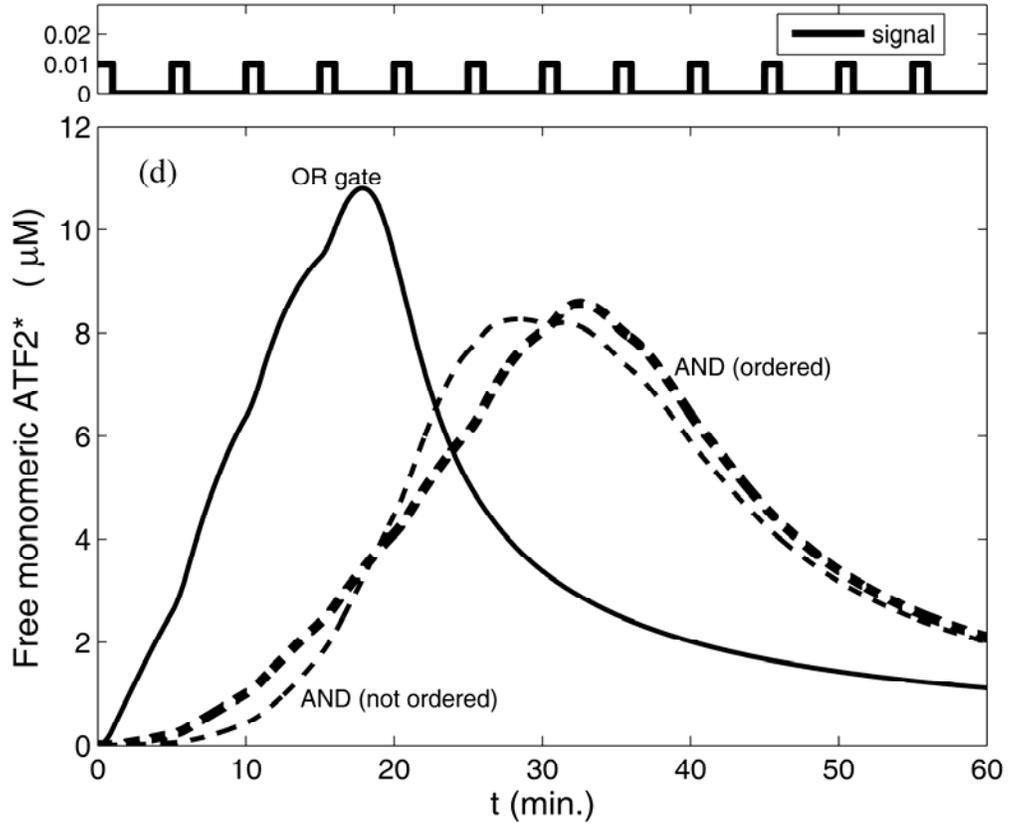

**Fig. 2e**

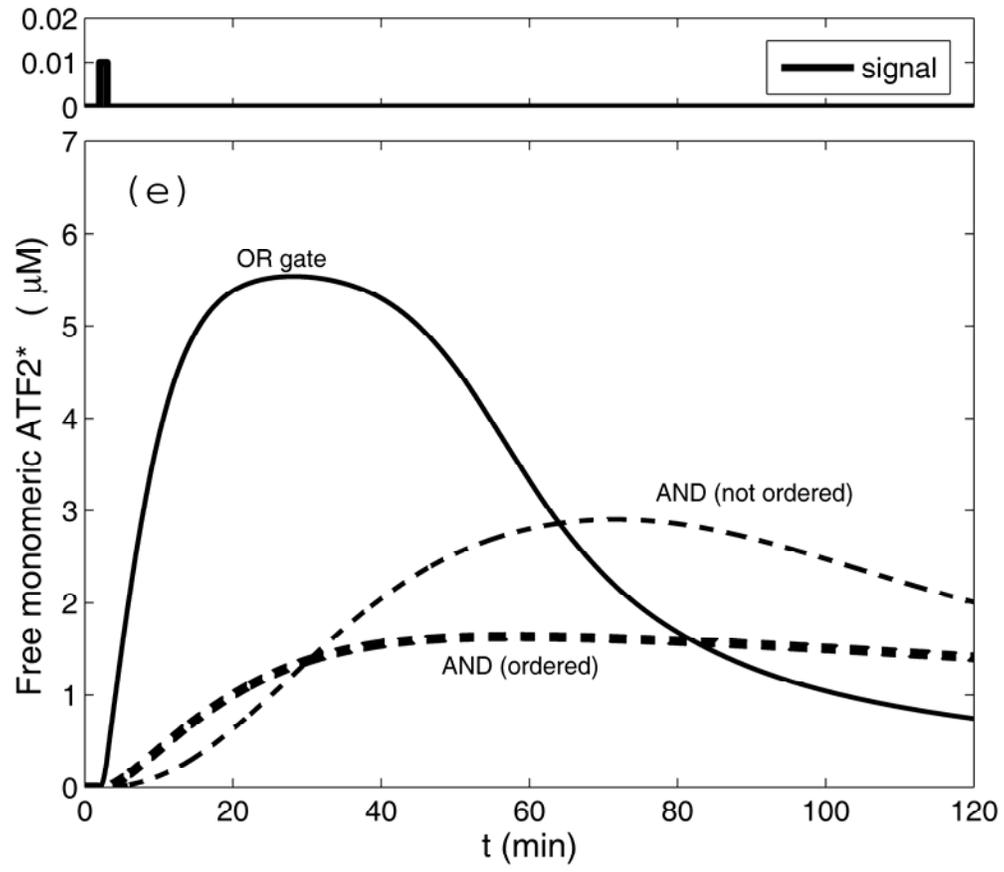

**Fig. 3a**

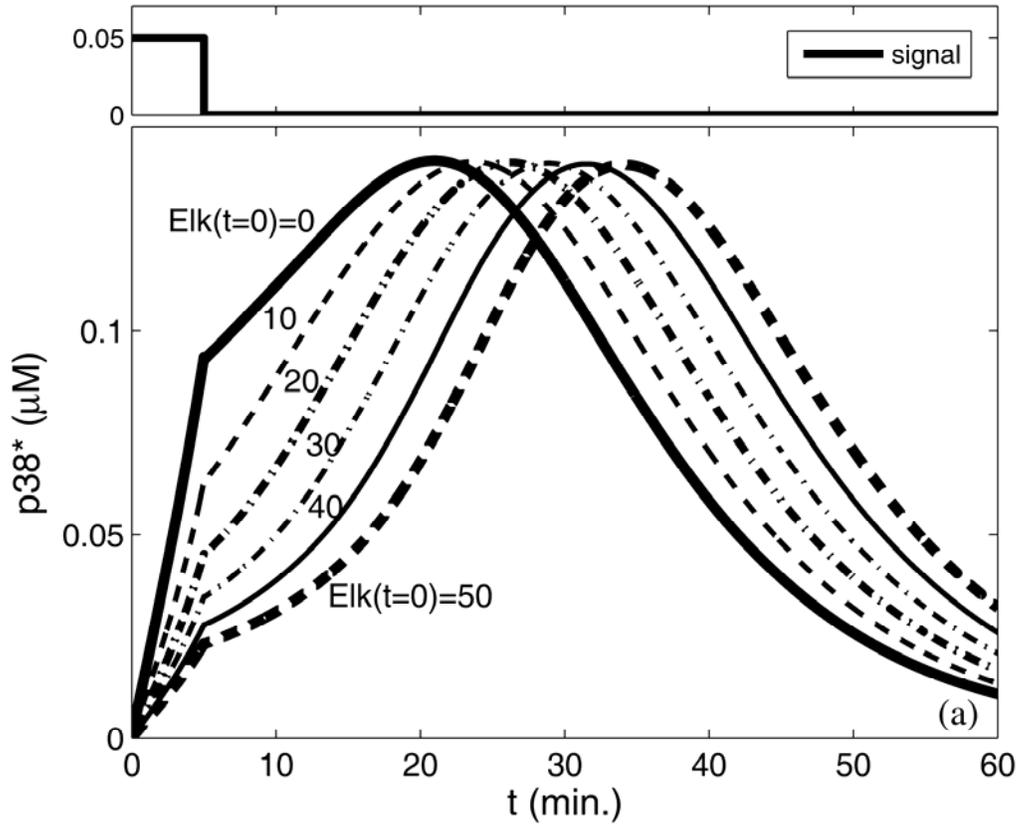

**Fig. 3b**

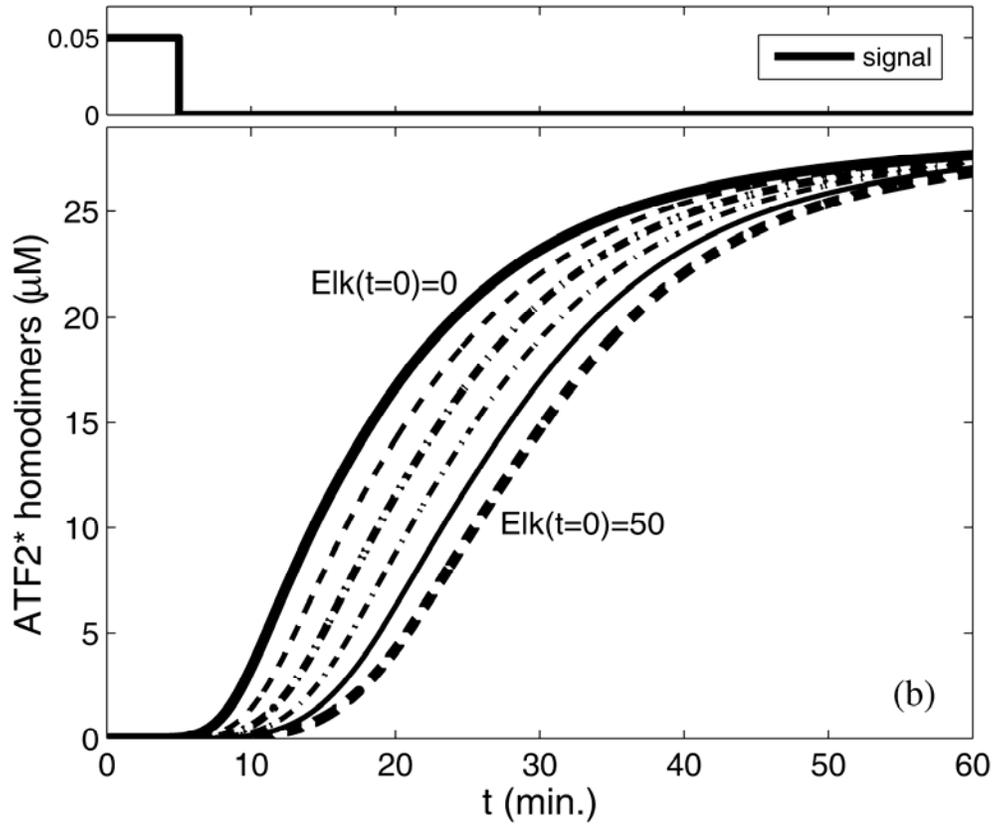

**Fig. 4a**

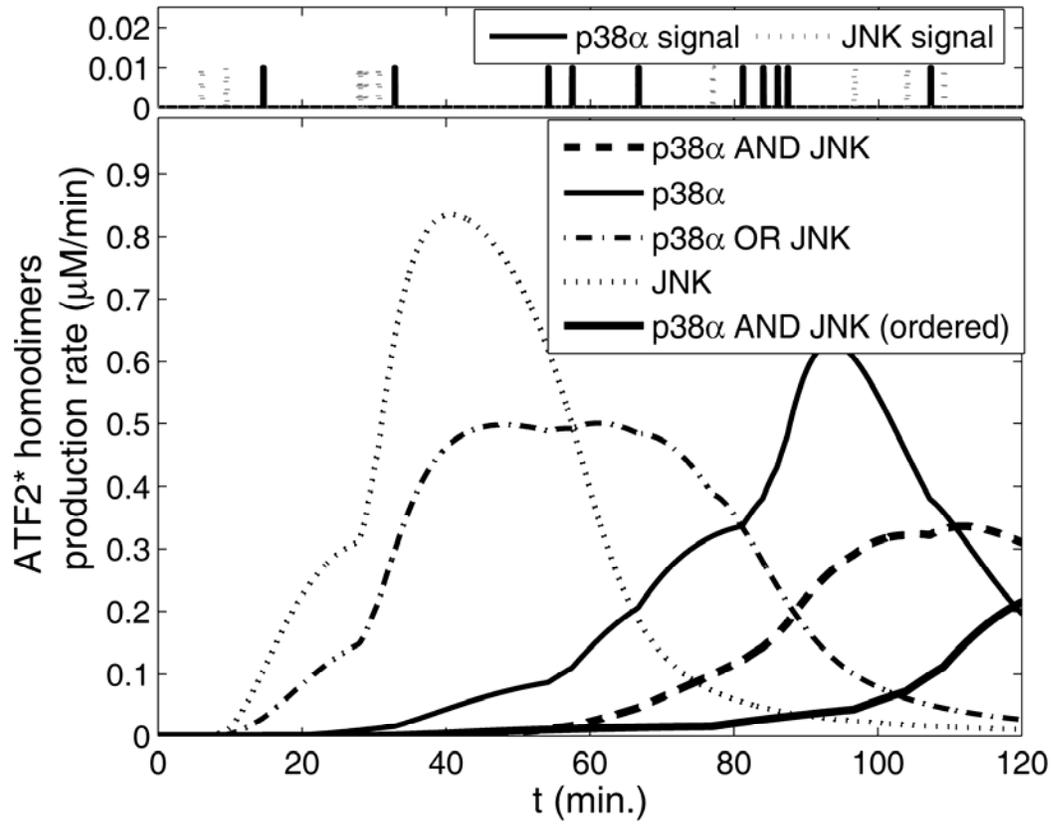

**Fig. 4b**

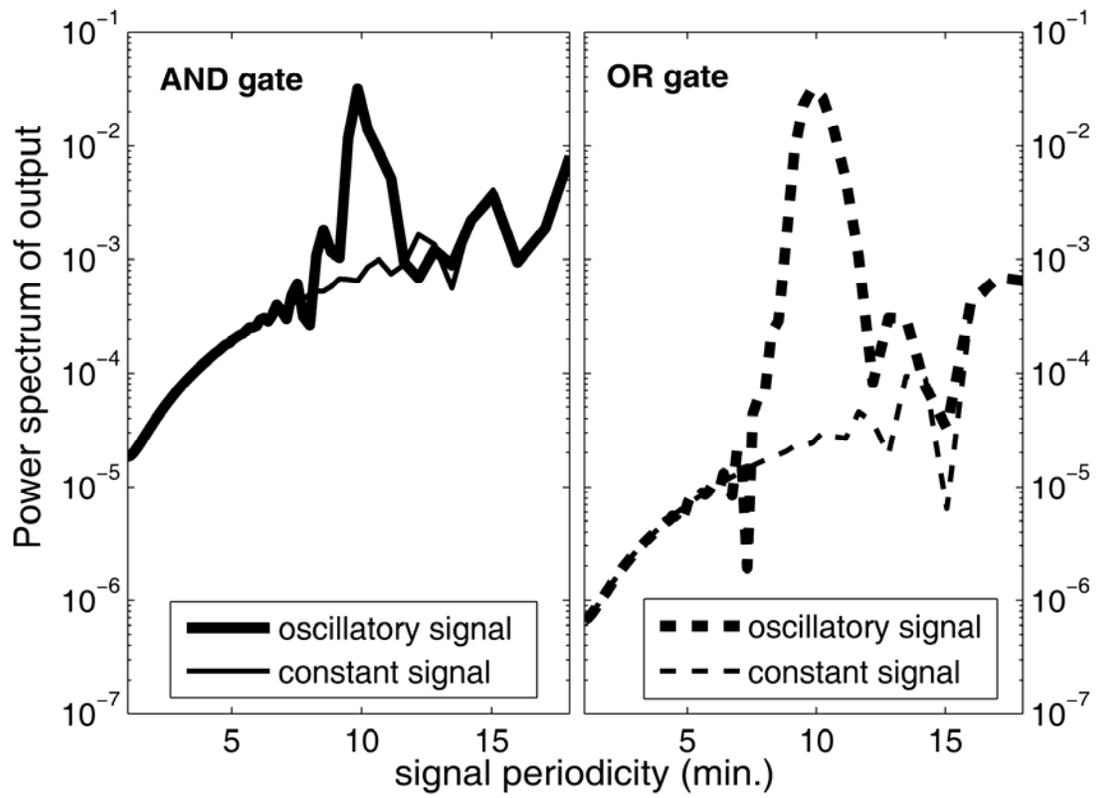

**Fig. 4c**

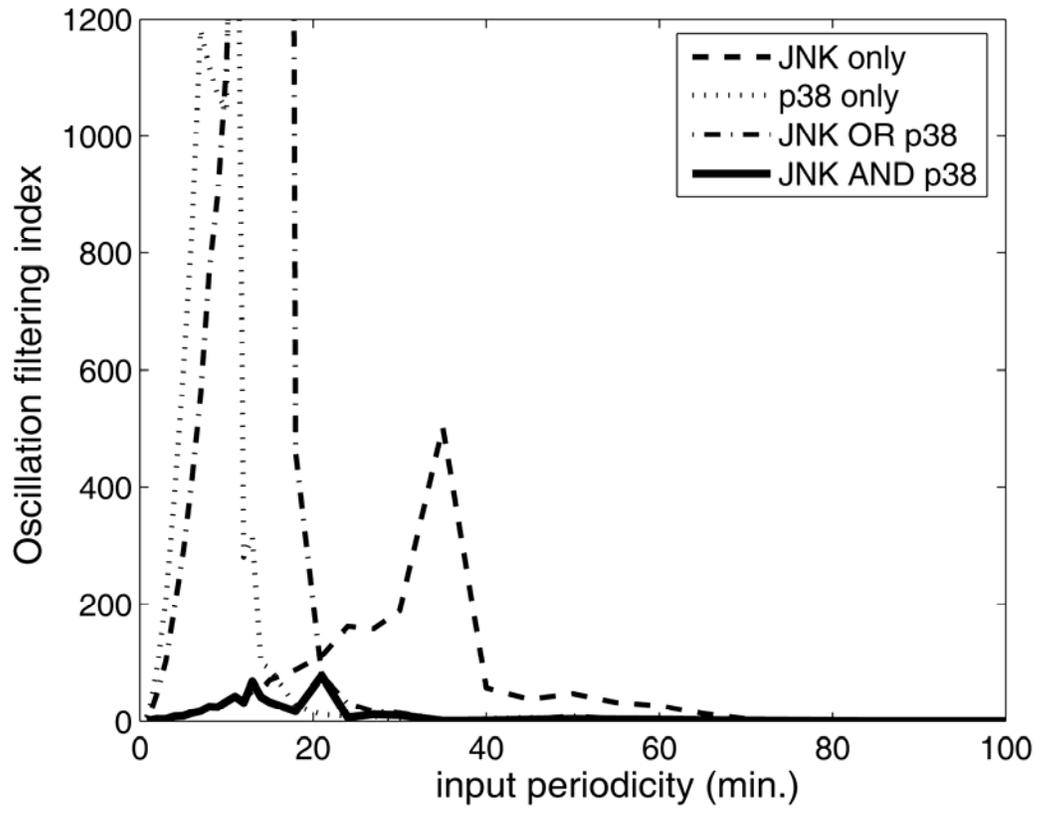

**Fig. 5**

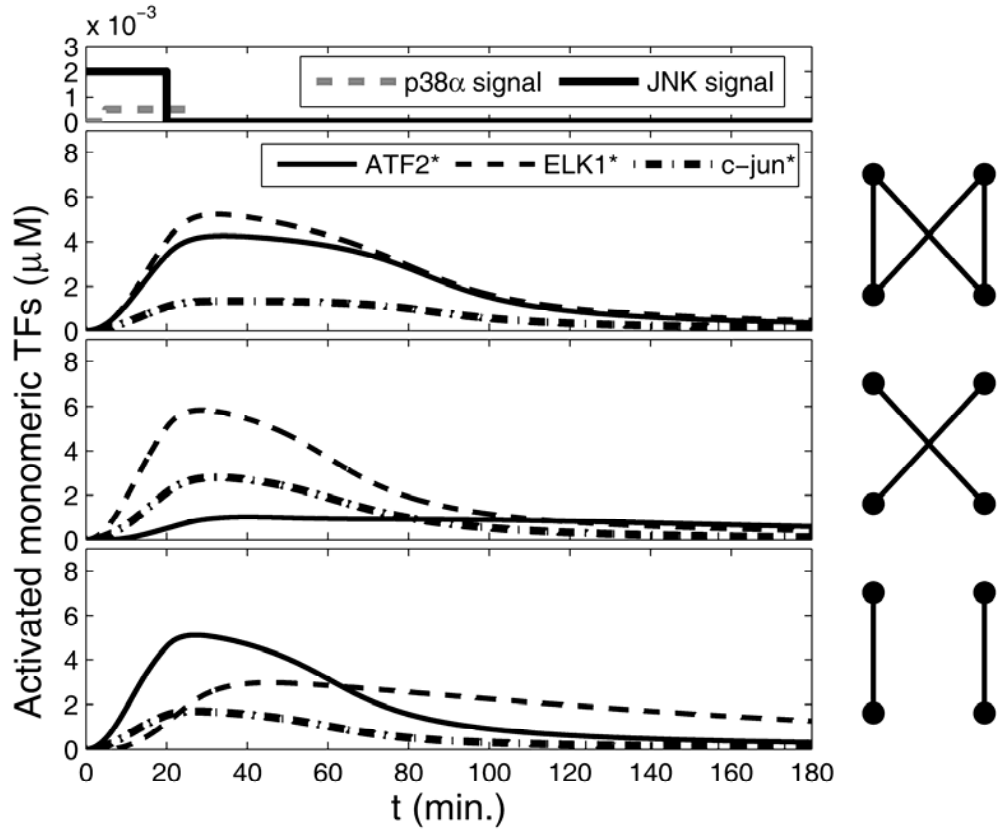

**Fig. 6**

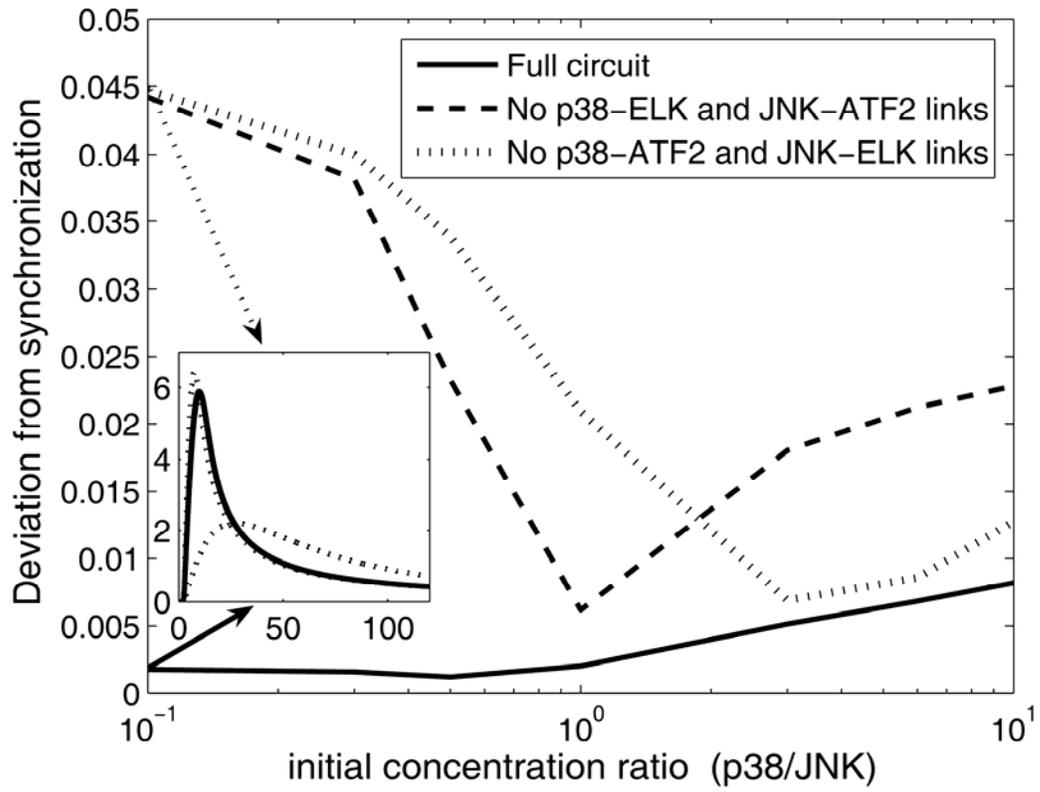

**Fig. 7**

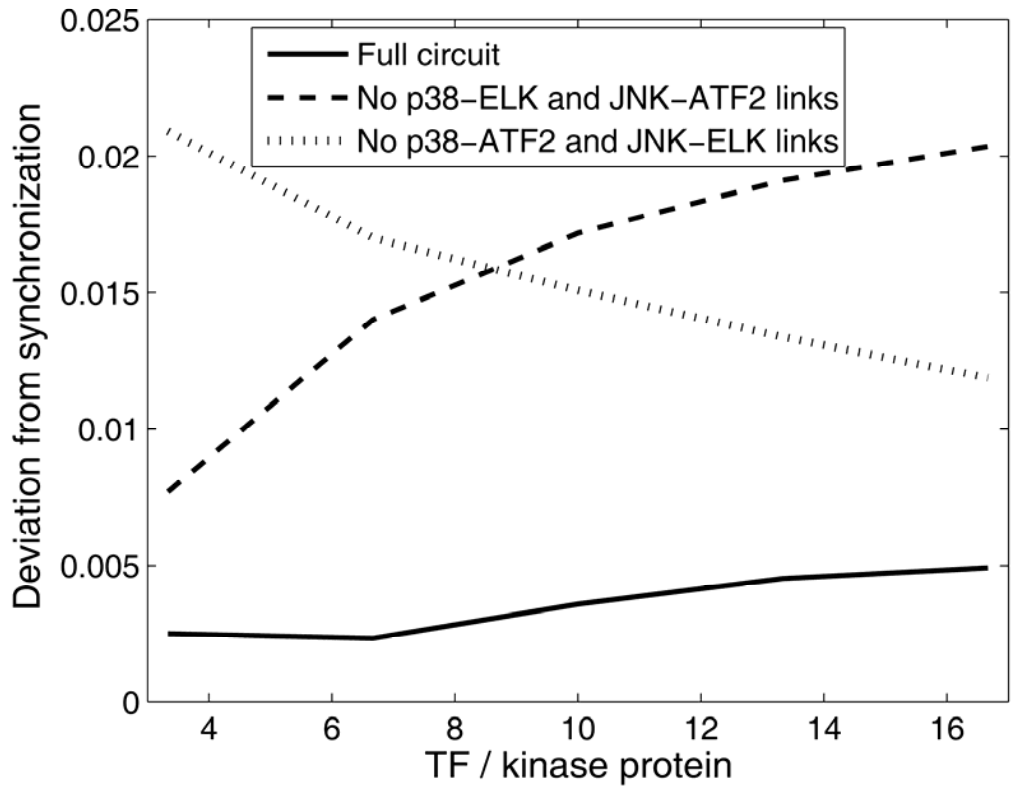

**Fig. 8**

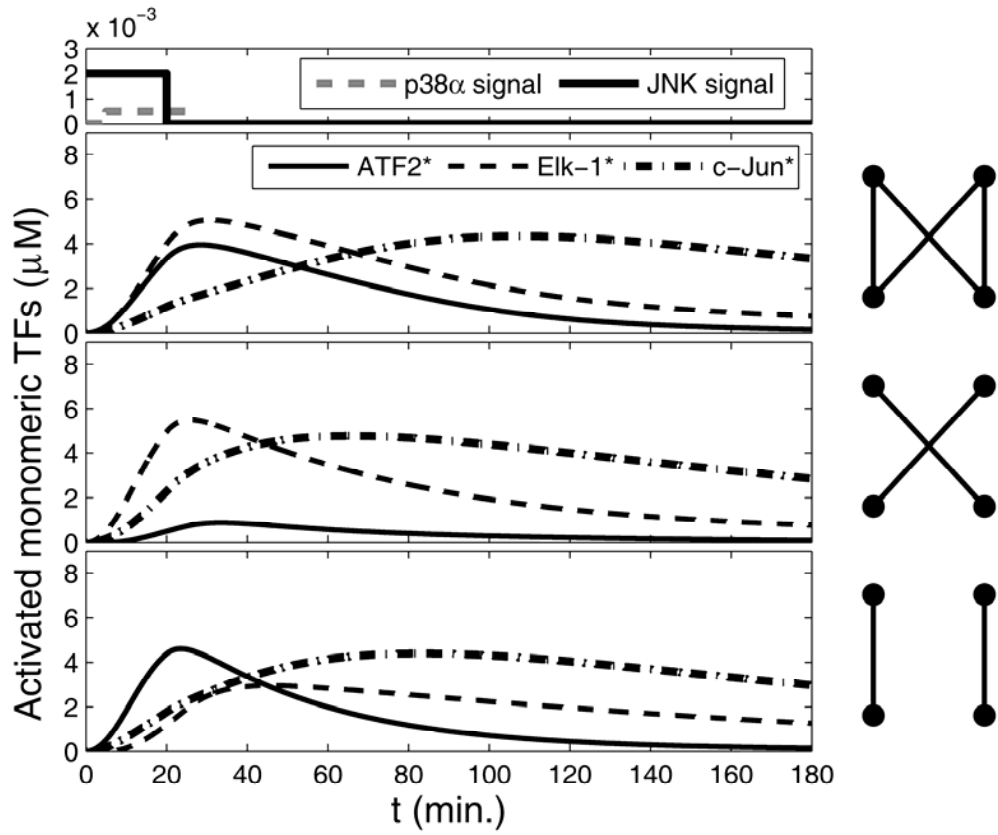

**Fig. 9**

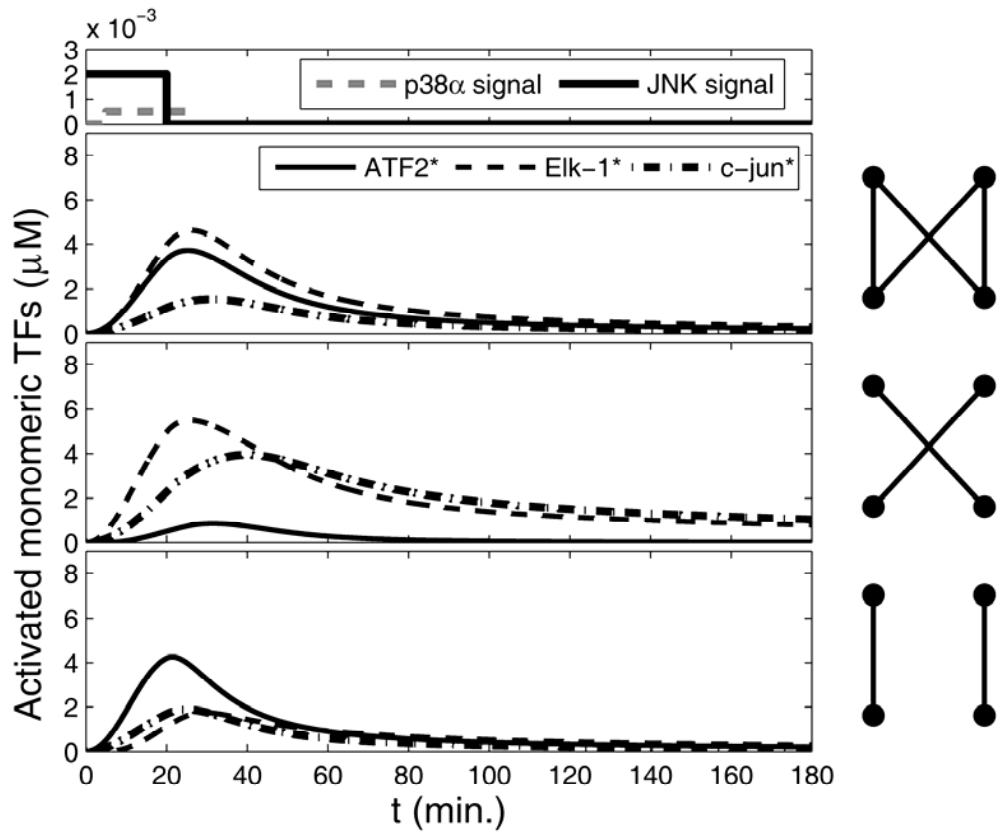

**Fig. 10**

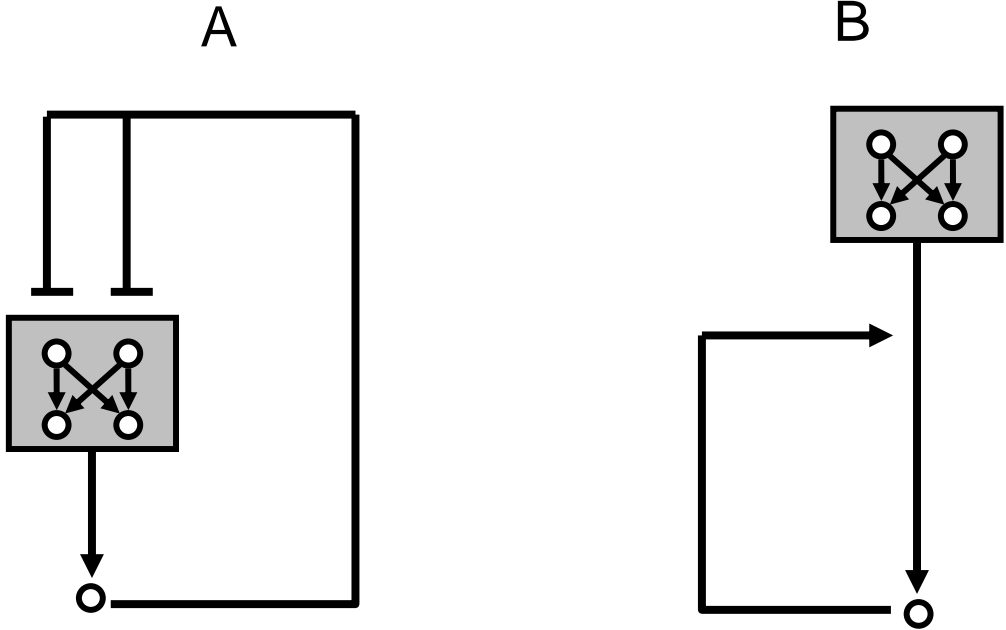

**Fig. 11**

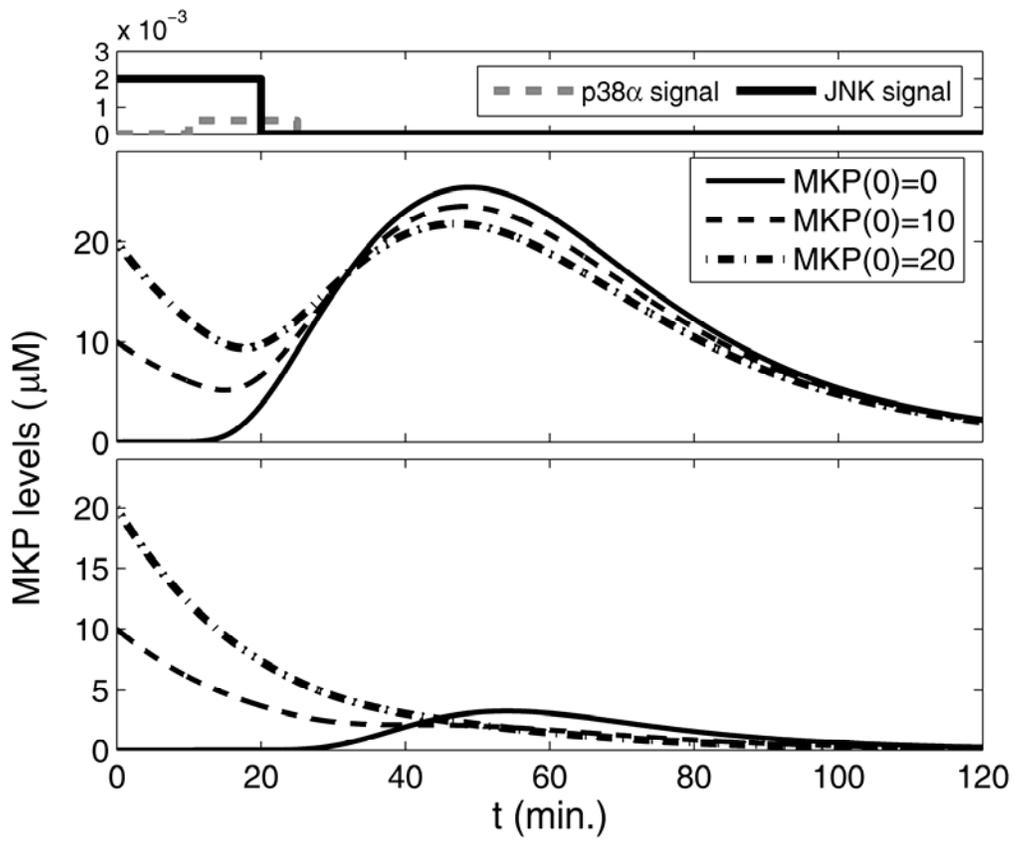

The following files are attached to the paper
"Functions of Bifans in Context of Multiple Regulatory Motifs in Signaling Networks":

`bifan1.M, bifan2.M` – These are the framework files, containing all required rates and information about required simulations (initial conditions, time span etc.). Using the for loops, it is possible to run these files once and get results for several different simulations. These files don't get any input and don't return any explicit output. However they create in the base workspace variables with simulations' results. In case of multiple simulations (e.g. with different initial conditions), the variables are enumerated accordingly. The file bifan1 direct to the OR gate configuration and bifan2 to the AND gate.

`and.M, and_order.M, or.M` – These file contain the call to the numerical solver and the equations of each of the bifan configurations. Return as output a vector 't' with time points, and a matrix 'c' of all concentrations at those time point.

`signalA.M, signalB.M` – The input signal of p38 and JNK respectively. Here we can put any function of time.

```matlab
%%%%%%%%%%%%%%%%%%%%%%%%%%%%%%%%%%%%%%%%%%%%%%%%%%%%%%%%%%%%%%%%%%%%%%%%%%%%%%%
%                                                                             %
% File name: bifan1.m                                                         %
% Attached to the paper:                                                      %
% Functions of Bifans in Context of Multiple Regulatory Motifs in  Signaling Networks  %
%                                                                             %
%%%%%%%%%%%%%%%%%%%%%%%%%%%%%%%%%%%%%%%%%%%%%%%%%%%%%%%%%%%%%%%%%%%%%%%%%%%%%%%

%  Parameters for OR gate simulation

global maxtime;

% variables:
% A – p38
% B – JNK
% C – ATF2
% D – ELK
% E – C-Jun
% M – MKP1
for ver=0:2 % version of cross linking
    for m=0:2 % version of circuit – with/without feedback loops

        k.init=[...These are the initial conditions. (In comment: variable name and
biological meaning)
            3   ...A         p38
            0   ...As        p38*
            3   ...B         JNK
            0   ...Bs        JNK*
            30  ...C          ATF2
            0   ...AsC       p38*:ATF2 complex
            0   ...BsC       JNK*:ATF2 complex
            0   ...Cs        ATF2*
            30  ...D          ELK
            0   ...AsD       p38*:ELK complex
            0   ...BsD       JNK*:ELK complex
            0   ...Ds        ELK*
            10  ...E          c-jun
            0   ...BsE       JNK*:c-jun complex
            0   ...Es        c-jun*
            0   ...CE        ATF2:c-jun complex
            0   ...E2        c-jun*:c-jun* complex
            0   ...M         MKP1
            0   ...C2promM   ATF2:ATF2 bound to MKP1 gene
            0   ...CEpromM   ATF2:c-jun bound to MKP1 gene
            0   ...CEpromE   ATF2:c-jun bound to c-june gene
            0   ...E2promE   c-jun:c-jun bound to c-june gene
            0];  % C2       ATF2:ATF2 complex
        %%%%%%%%%%%%%%%%%%%%%%%%%%%%%%%%%%%%%%%%%%%%%%%%%%%%%%%%%
        % Reaction rates:
        % (To eliminate links from the circuit,
        %  set reaction rate to zero)
        %%%%%%%%%%%%%%%%%%%%%%%%%%%%%%%%%%%%%%%%%%%%%%%%%%%%%%%%%

        k.A_As=0; %  signal A
        k.As_A=0.1;
        k.As_C_AsC=3.9*(ver~=2);
        k.AsC_As_Cs=4.8;
        k.AsC_As_C=19.2;
        k.As_D_AsD=16.02*(ver~=1);
        k.AsD_As_Ds=8.75;
        k.AsD_As_D=35;

        k.B_Bs=0;%  signal B
        k.Bs_B=0.1;
        k.Bs_C_BsC=20*(ver~=1);
        k.BsC_Bs_Cs=10;
        k.BsC_Bs_C=40;
        k.Bs_D_BsD=020*(ver~=2);
        k.BsD_Bs_Ds=10;
        k.BsD_Bs_D=40;
```

```matlab
            k.Bs_E_BsE=20;
            k.BsE_Bs_Es=10;
            k.BsE_Bs_E=40;

            k.C_trans=0.0;
            k.D_trans=0.0;
            k.E_trans=1*(m>0);
            k.M_trans=3*(m>1);
            k.M_deg=0.05;
            k.E_deg=0.1*(m>0);
            k.E2_deg=0.2;
            k.CE_deg=0.2;
            k.C2_deg=0.2;
            k.D_deg=0.0;
            k.C_deg=0.0;

            k.Cs_C=0;
            k.Ds_D=0;
            k.Es_E=0;

            k.promMbinding=5*(m>1);
            k.promMunbind=01.0;
            k.promEbinding=20*(m>0);
            k.promEunbind=10;
            k.Mdephos=0.10;

            k.Cs_Cs_C2=.02;
            k.Cs_Es_CE=.02;
            k.Cs_Ds_CD=0.0;
            k.Ds_Ds_D2=.02;
            k.Es_Es_E2=.02;

            k.Ds_Es_DE=0.0;

            % ODEs + Numerical solver:
            [t_or,c_or]=or(maxtime,k);

            % Output:
            % variables are created in base workspace.

            assignin('caller',['t_v' num2str(ver) '_' num2str(m)],t_or);
            assignin('caller',['c_v' num2str(ver) '_' num2str(m)],c_or);

    end
end
```

```matlab
%%%%%%%%%%%%%%%%%%%%%%%%%%%%%%%%%%%%%%%%%%%%%%%%%%%%%%%%%%%%%%%%%%%%%%%%%%%%%
%                                                                           %
% File name: bifan2.m                                                       %
% Attached to the paper:                                                    %
% Functions of Bifans in Context of Multiple Regulatory Motifs in  Signaling Networks  %
%                                                                           %
%%%%%%%%%%%%%%%%%%%%%%%%%%%%%%%%%%%%%%%%%%%%%%%%%%%%%%%%%%%%%%%%%%%%%%%%%%%%%

%  Parameters for AND gate simulation
global maxtime;

% variables:
% A - p38
% B - JNK
% C - ATF2
% D - ELK
% E - C-Jun
% M - MKP1

for m=3:3 % use this loop to have several sets of initial condition, as function of m

    k.init=[...These are the initial conditions.
        10 ...A          p38
        0 ...As          p38*
        10 ...B          JNK
        0 ...Bs          JNK*
        30 ...C          ATF2
        0 ...AsC         p38*:ATF2 complex
        0 ...BsC         JNK*:ATF2 complex
        0 ...Csa         ATF2*
        0 ...Csb
        0 ...AsCs
        0 ...BsCs
        0 ...Css
        10*m ...D        ELK
        0 ...AsD         p38*:ELK complex
        0 ...BsD         JNK*:ELK complex
        0 ...Dsa         ELK*
        0 ...Dsb
        0 ...AsDs
        0 ...BsDs
        0 ...Dss
        0]; % C2         ATF2:ATF2 complex
    %%%%%%%%%%%%%%%%%%%%%%%%%%%%%%%%%%%%%%%%%%%%%%%%%%%%%%%%%
    % Reaction rates:
    % (To eliminate links from the circuit,
    %  set reaction rate to zero)
    %%%%%%%%%%%%%%%%%%%%%%%%%%%%%%%%%%%%%%%%%%%%%%%%%%%%%%%%%

    k.A_As=0; %   signal A
    k.As_A=0.1;
    k.As_C_AsC=3.9;
    k.AsC_As_Cs=4.8;
    k.AsC_As_C=19.2;
    k.As_D_AsD=16.02;
    k.AsD_As_Ds=8.75;
    k.AsD_As_D=35;

    k.B_Bs=0;%   signal B
    k.Bs_B=0.1;
    k.Bs_C_BsC=20;
    k.BsC_Bs_Cs=10;
    k.BsC_Bs_C=40;
    k.Bs_D_BsD=20;
    k.BsD_Bs_Ds=10;
    k.BsD_Bs_D=40;
    k.Bs_E_BsE=20;
    k.BsE_Bs_Es=10;
    k.BsE_Bs_E=40;
```

```
        k.C_trans=0.0;
        k.D_trans=0.0;
        k.E_trans=1;
        k.M_trans=3;
        k.M_deg=0.05;
        k.E_deg=0.1;
        k.E2_deg=0.2;
        k.CE_deg=0.2;
        k.C2_deg=0.2;
        k.D_deg=0.0;
        k.C_deg=0.0;

        k.Cs_C=0.0;
        k.Ds_D=0.0;
        k.Es_E=0.0;

        k.promMbinding=5;
        k.promMunbind=1.0;
        k.promEbinding=20;
        k.promEunbind=10;
        k.Mdephos=0.1;

        k.Cs_Cs_C2=0.02;
        k.Cs_Es_CE=0.02;
        k.Cs_Ds_CD=0.0;
        k.Ds_Ds_D2=0.02;
        k.Es_Es_E2=0.02;

        k.Ds_Es_DE=0.0;

        % ODEs + Numerical solver:
        [t_and,c_and]=and(maxtime,k);

        % For the ordered version, replace previous line with the following:
        % [t_and,c_and]=and_order(maxtime,k);
        % and don't forget to update required rates, if necessary.

        % Output:
        % variables are created in base workspace.
        assignin('base',['t_and_' num2str(m)],t_and);
        assignin('base',['c_and_' num2str(m)],c_and);
end
```

```matlab
%%%%%%%%%%%%%%%%%%%%%%%%%%%%%%%%%%%%%%%%%%%%%%%%%%%%%%%%%%%%%%%%%%%%%%%%%%%
%                                                                         %
% File name: and.m                                                        %
% Attached to the paper:                                                  %
% Functions of Bifans in Context of Multiple Regulatory Motifs in  Signaling Networks  %
%                                                                         %
%%%%%%%%%%%%%%%%%%%%%%%%%%%%%%%%%%%%%%%%%%%%%%%%%%%%%%%%%%%%%%%%%%%%%%%%%%%

function [t y]=and(maxtime,k)

% AND gate
options=odeset('RelTol',1e-6,'AbsTol',1e-6,'MaxStep',0.15);

% A As B Bs C AC BC Cs D AD BD Ds E BE Es
init_cond=k.init;

[t,y]=ode45(@func,[0,maxtime],init_cond,options,k);

%--------------------------------------------------

function ddt=func(t,N,k)

A=N(1);
As=N(2);
B=N(3);
Bs=N(4);
C=N(5);
AsC=N(6);
BsC=N(7);
Csa=N(8);
Csb=N(9);
AsCs=N(10);
BsCs=N(11);
Css=N(12);
D=N(13);
AsD=N(14);
BsD=N(15);
Dsa=N(16);
Dsb=N(17);
AsDs=N(18);
BsDs=N(19);
Dss=N(20);
C2=N(21);

% These are the equations for non-ordered AND gate.

%A
ddt(1)=-signalA(t)*A+k.As_A*As;
%As
ddt(2)=signalA(t)*(A)-k.As_A*As...
    -k.As_C_AsC*As*C+(k.AsC_As_C+k.AsC_As_Cs)*AsC...
    -k.As_D_AsD*As*D+(k.AsD_As_D+k.AsD_As_Ds)*AsD...
    -k.As_C_AsC*As*Csb+(k.AsC_As_C+k.AsC_As_Cs)*AsCs...
    -k.As_D_AsD*As*Dsb+(k.AsD_As_D+k.AsD_As_Ds)*AsDs;
%B
ddt(3)=-signalB(t)*(B)+k.Bs_B*Bs;
%Bs
ddt(4)=signalB(t)*(B)-k.Bs_B*Bs...
    -k.Bs_C_BsC*Bs*C+(k.BsC_Bs_C+k.BsC_Bs_Cs)*BsC...
    -k.Bs_D_BsD*Bs*D+(k.BsD_Bs_D+k.BsD_Bs_Ds)*BsD...
    -k.Bs_C_BsC*Bs*Csa+(k.BsC_Bs_C+k.BsC_Bs_Cs)*BsCs...
    -k.Bs_D_BsD*Bs*Dsa+(k.BsD_Bs_D+k.BsD_Bs_Ds)*BsDs;
%C
ddt(5)=k.C_trans-k.As_C_AsC*As*C+k.AsC_As_C*AsC-
k.Bs_C_BsC*Bs*C+k.BsC_Bs_C*BsC+k.Cs_C*(Csa+Csb)-k.C_deg*C;
%AsC
```

```
ddt(6)=k.As_C_AsC*As*C-(k.AsC_As_C+k.AsC_As_Cs)*AsC;
%BsC
ddt(7)=k.Bs_C_BsC*Bs*C-(k.BsC_Bs_C+k.BsC_Bs_Cs)*BsC;
%Csa
ddt(8)=k.AsC_As_Cs*AsC-k.Bs_C_BsC*Bs*Csa+k.BsC_Bs_C*BsCs-k.Cs_C*Csa;
%Csb
ddt(9)=k.BsC_Bs_Cs*BsC-k.As_C_AsC*As*Csb+k.AsC_As_C*AsCs-k.Cs_C*Csb;
%AsCs
ddt(10)=k.As_C_AsC*As*Csb-(k.AsC_As_C+k.AsC_As_Cs)*AsCs;
%BsCs
ddt(11)=k.Bs_C_BsC*Bs*Csa-(k.BsC_Bs_C+k.BsC_Bs_Cs)*BsCs;
%Css
ddt(12)=k.AsC_As_Cs*AsCs+k.BsC_Bs_Cs*BsCs-k.Cs_Cs_C2*Css^2-k.Cs_Ds_CD*Css*Dss-k.Cs_C*Css;
%D
ddt(13)=k.D_trans-k.As_D_AsD*As*D+k.AsD_As_D*AsD-
k.Bs_D_BsD*Bs*D+k.BsD_Bs_D*BsD+k.Ds_D*(Dsa+Dsb)-k.D_deg*D;
%AsD
ddt(14)=k.As_D_AsD*As*D-(k.AsD_As_D+k.AsD_As_Ds)*AsD;
%BsD
ddt(15)=k.Bs_D_BsD*Bs*D-(k.BsD_Bs_D+k.BsD_Bs_Ds)*BsD;
%Dsa
ddt(16)=k.AsD_As_Ds*AsD-k.Bs_D_BsD*Bs*Dsa+k.BsD_Bs_D*BsDs-k.Ds_D*Dsa;
%Dsb
ddt(17)=k.BsD_Bs_Ds*BsD-k.As_D_AsD*As*Dsb+k.AsD_As_D*AsDs-k.Ds_D*Dsb;
%AsDs
ddt(18)=k.As_D_AsD*As*Dsb-(k.AsD_As_D+k.AsD_As_Ds)*AsDs;
%BsDs
ddt(19)=k.Bs_D_BsD*Bs*Dsa-(k.BsD_Bs_D+k.BsD_Bs_Ds)*BsDs;
%Dss
ddt(20)=k.AsD_As_Ds*AsDs+k.BsD_Bs_Ds*BsDs-k.Ds_Ds_D2*Dss^2-k.Cs_Ds_CD*Css*Dss-k.Ds_D*Dss;
% C2
ddt(21)=k.Cs_Cs_C2*Css^2-k.C2_deg*C2;

ddt=ddt';
```

```matlab
%%%%%%%%%%%%%%%%%%%%%%%%%%%%%%%%%%%%%%%%%%%%%%%%%%%%%%%%%%%%%%%%%%%%%%%%%%%
%                                                                         %
% File name: and_order.m
%
% Attached to the paper:                                                  %
% Functions of Bifans in Context of Multiple Regulatory Motifs in  Signaling Networks   %
%                                                                         %
%%%%%%%%%%%%%%%%%%%%%%%%%%%%%%%%%%%%%%%%%%%%%%%%%%%%%%%%%%%%%%%%%%%%%%%%%%%

function [t y]=and_ord(maxtime,k)

options=odeset('RelTol',1e-8,'AbsTol',1e-9,'MaxStep',0.15);

% A As B Bs C AC BC Cs D AD BD Ds E BE Es
init_cond=k.init;

[t,y]=ode45(@func,[0,maxtime],init_cond,options,k);

%-------------------------------------------------

function ddt=func(t,N,k)

A=N(1);
As=N(2);
B=N(3);
Bs=N(4);
C=N(5);
AsC=N(6);
BsC=N(7);
Csa=N(8);
Csb=N(9);
AsCs=N(10);
BsCs=N(11);
Css=N(12);
D=N(13);
AsD=N(14);
BsD=N(15);
Dsa=N(16);
Dsb=N(17);
AsDs=N(18);
BsDs=N(19);
Dss=N(20);
C2=N(21);

% These are the equations for ordered AND gate.

%A
ddt(1)=-(signalA(t)*(A))+k.As_A*As;
%As
ddt(2)=signalA(t)*(A)-k.As_A*As...
    -k.As_C_AsC*As*C+(k.AsC_As_C+k.AsC_As_Cs)*AsC...
    -k.As_D_AsD*As*D+(k.AsD_As_D+k.AsD_As_Ds)*AsD;
%B
ddt(3)=-signalB(t)*(B)+k.Bs_B*Bs;
%Bs
ddt(4)=signalB(t)*(B)-k.Bs_B*Bs...
    -k.Bs_C_BsC*Bs*Csa+(k.BsC_Bs_C+k.BsC_Bs_Cs)*BsCs...
    -k.Bs_D_BsD*Bs*Dsa+(k.BsD_Bs_D+k.BsD_Bs_Ds)*BsDs;
%C
ddt(5)=k.C_trans-k.As_C_AsC*As*C+k.AsC_As_C*AsC+k.Cs_C*(Csa)-k.C_deg*C;
%AsC
ddt(6)=k.As_C_AsC*As*C-(k.AsC_As_C+k.AsC_As_Cs)*AsC;
%BsC
ddt(7)=0;
%Csa
ddt(8)=k.AsC_As_Cs*AsC-k.Bs_C_BsC*Bs*Csa+k.BsC_Bs_C*BsCs-k.Cs_C*Csa+k.Cs_C*Css;
%Csb
ddt(9)=0;
%AsCs
```

```
    ddt(10)=0;
    %BsCs
    ddt(11)=k.Bs_C_BsC*Bs*Csa-(k.BsC_Bs_C+k.BsC_Bs_Cs)*BsCs;
    %Css
    ddt(12)=k.BsC_Bs_Cs*BsCs-k.Cs_Cs_C2*Css^2-k.Cs_Ds_CD*Css*Dss-k.Cs_C*Css;
    %D
    ddt(13)=k.D_trans-k.As_D_AsD*As*D+k.AsD_As_D*AsD+k.Ds_D*(Dsa+Dsb)-k.D_deg*D;
    %AsD
    ddt(14)=k.As_D_AsD*As*D-(k.AsD_As_D+k.AsD_As_Ds)*AsD;
    %BsD
    ddt(15)=0;
    %Dsa
    ddt(16)=k.AsD_As_Ds*AsD-k.Bs_D_BsD*Bs*Dsa+k.BsD_Bs_D*BsDs-k.Ds_D*Dsa+k.Ds_D*Dss;
    %Dsb
    ddt(17)=0;
    %AsDs
    ddt(18)=0;
    %BsDs
    ddt(19)=k.Bs_D_BsD*Bs*Dsa-(k.BsD_Bs_D+k.BsD_Bs_Ds)*BsDs;
    %Dss
    ddt(20)=k.BsD_Bs_Ds*BsDs-k.Ds_Ds_D2*Dss^2-k.Cs_Ds_CD*Css*Dss-k.Ds_D*Dss;
    % C2
    ddt(21)=k.Cs_Cs_C2*Css^2-k.C2_deg*C2;

    ddt=ddt';
```

```matlab
%%%%%%%%%%%%%%%%%%%%%%%%%%%%%%%%%%%%%%%%%%%%%%%%%%%%%%%%%%%%%%%%%%%%%%%%%%%
%                                                                         %
% File name: or.m                                                         %
% Attached to the paper:                                                  %
% Functions of Bifans in Context of Multiple Regulatory Motifs in  Signaling Networks  %
%                                                                         %
%%%%%%%%%%%%%%%%%%%%%%%%%%%%%%%%%%%%%%%%%%%%%%%%%%%%%%%%%%%%%%%%%%%%%%%%%%%

function [t y]=or (maxtime,k)

% OR gate
options=odeset('RelTol',1e-6,'AbsTol',1e-6,'MaxStep',0.15);

% A As B Bs C AC BC Cs D AD BD Ds E BE Es
init_cond=k.init;

[t,y]=ode45(@func,[0,maxtime],init_cond,options,k);

%--------------------------------------------------

function ddt=func(t,N,k)

A=N(1);
As=N(2);
B=N(3);
Bs=N(4);
C=N(5);
AsC=N(6);
BsC=N(7);
Cs=N(8);
D=N(9);
AsD=N(10);
BsD=N(11);
Ds=N(12);
E=N(13);
BsE=N(14);
Es=N(15);
CE=N(16);
E2=N(17);
M=N(18);
C2promM=N(19);
CEpromM=N(20);
CEpromE=N(21);
E2promE=N(22);
C2=N(23);

%A
ddt(1)=-signalA(t)*A+k.As_A*As+k.Mdephos*M*As;
%As
ddt(2)=signalA(t)*(A)-k.As_A*As...
    -k.As_C_AsC*As*C+(k.AsC_As_C+k.AsC_As_Cs)*AsC...
    -k.As_D_AsD*As*D+(k.AsD_As_D+k.AsD_As_Ds)*AsD...
    -k.Mdephos*M*As;
%B
ddt(3)=-signalB(t)*(B)+k.Bs_B*Bs+k.Mdephos*M*Bs;
%Bs
ddt(4)=signalB(t)*(B)-k.Bs_B*Bs...
    -k.Bs_C_BsC*Bs*C+(k.BsC_Bs_C+k.BsC_Bs_Cs)*BsC...
    -k.Bs_D_BsD*Bs*D+(k.BsD_Bs_D+k.BsD_Bs_Ds)*BsD...
    -k.Bs_E_BsE*Bs*E+(k.BsE_Bs_E+k.BsE_Bs_Es)*BsE...
    -k.Mdephos*M*Bs;
%C
ddt(5)=k.C_trans-k.As_C_AsC*As*C+k.AsC_As_C*AsC-k.Bs_C_BsC*Bs*C+k.BsC_Bs_C*BsC+k.Cs_C*Cs-k.C_deg*C;
%AsC
ddt(6)=k.As_C_AsC*As*C-(k.AsC_As_C+k.AsC_As_Cs)*AsC;
%BsC
ddt(7)=k.Bs_C_BsC*Bs*C-(k.BsC_Bs_C+k.BsC_Bs_Cs)*BsC;
```

```
%Cs
ddt(8)=k.AsC_As_Cs*AsC+k.BsC_Bs_Cs*BsC-k.Cs_Cs_C2*Cs^2-k.Cs_Ds_CD*Cs*Ds-k.Cs_Es_CE*Cs*Es-k.Cs_C*Cs;
%D
ddt(9)=k.D_trans-k.As_D_AsD*As*D+k.AsD_As_D*AsD-k.Bs_D_BsD*Bs*D+k.BsD_Bs_D*BsD+k.Ds_D*Ds-k.D_deg*D;
%AsD
ddt(10)=k.As_D_AsD*As*D-(k.AsD_As_D+k.AsD_As_Ds)*AsD;
%BsD
ddt(11)=k.Bs_D_BsD*Bs*D-(k.BsD_Bs_D+k.BsD_Bs_Ds)*BsD;
%Ds
ddt(12)=k.AsD_As_Ds*AsD+k.BsD_Bs_Ds*BsD-k.Ds_Ds_D2*Ds^2-k.Ds_Es_DE*Es*Ds-k.Ds_D*Ds-k.Cs_Ds_CD*Cs*Ds;
%E
ddt(13)=-k.Bs_E_BsE*Bs*E+k.BsE_Bs_E*BsE+k.Es_E*Es-k.E_deg*E+k.E_trans*(1.0+(CEpromE+E2promE)/.001);
%BsE
ddt(14)=k.Bs_E_BsE*Bs*E-(k.BsE_Bs_E+k.BsE_Bs_Es)*BsE;
%Es
ddt(15)=k.BsE_Bs_Es*BsE-k.Es_Es_E2*Es^2-k.Cs_Es_CE*Cs*Es-k.Ds_Es_DE*Es*Ds-k.Es_E*Es;
%CE
ddt(16)=k.Cs_Es_CE*Cs*Es-k.promMbinding*CE*(.001-CEpromM)+k.promMunbind*CEpromM-k.promEbinding*CE*(.001-CEpromE)+k.promEunbind*CEpromE-k.CE_deg*CE;
%E2
ddt(17)=k.Es_Es_E2*Es^2-k.promEbinding*E2*(.001-E2promE)+k.promEunbind*E2promE-k.E2_deg*E2;
%M
ddt(18)=-k.M_deg*M+k.M_trans*(C2promM*CEpromM)/.001^2;
% C2promM
ddt(19)=-k.promMunbind*C2promM+k.promMbinding*C2*(.001-C2promM);
% CEpromM
ddt(20)=k.promMbinding*CE*(.001-CEpromM)-k.promMunbind*CEpromM;
% CEpromE
ddt(21)=k.promEbinding*CE*(.001-CEpromE)-k.promEunbind*CEpromE;
% E2promE
ddt(22)=k.promEbinding*E2*(.001-E2promE)-k.promEunbind*E2promE;
% C2
ddt(23)=k.Cs_Cs_C2*Cs^2-k.promMbinding*C2*(.001-C2promM)+k.promMunbind*C2promM-k.C2_deg*C2;

ddt=ddt';
```

```
%%%%%%%%%%%%%%%%%%%%%%%%%%%%%%%%%%%%%%%%%%%%%%%%%%%%%%%%%%%%%%%%%%%%%%%%%%%%%%%
%                                                                             %
% File name: signalA.m                                                        %
% Attached to the paper:                                                      %
% Functions of Bifans in Context of Multiple Regulatory Motifs in  Signaling Networks  %
%                                                                             %
%%%%%%%%%%%%%%%%%%%%%%%%%%%%%%%%%%%%%%%%%%%%%%%%%%%%%%%%%%%%%%%%%%%%%%%%%%%%%%%

function s=signalA(t)

s=0.0;
if (t<25 & t>5)
    s=0.0005;
end
```

```
%%%%%%%%%%%%%%%%%%%%%%%%%%%%%%%%%%%%%%%%%%%%%%%%%%%%%%%%%%%%%%%%%%%%%%%%%%%%%%%
%                                                                             %
% File name: signalB.m                                                        %
% Attached to the paper:                                                      %
% Functions of Bifans in Context of Multiple Regulatory Motifs in  Signaling Networks  %
%                                                                             %
%%%%%%%%%%%%%%%%%%%%%%%%%%%%%%%%%%%%%%%%%%%%%%%%%%%%%%%%%%%%%%%%%%%%%%%%%%%%%%%

function s=signalB(t)
s=0;
if t<20
    s=0.002;
end
```